 \documentclass[aps,pra,twocolumn]{revtex4-1}
\usepackage{array}
\usepackage{upgreek}
\usepackage{epsfig}
\usepackage{tikz}
\usepackage{graphicx}
\usepackage{epstopdf}
\usepackage{fancybox}
\usepackage{makeidx}
\usepackage{mathrsfs}
\usepackage{appendix}
\usepackage{amsmath}
\usepackage{amsfonts}
\usepackage{natbib}
\usepackage[figuresright]{rotating}
\usepackage{floatpag}
\usepackage{natbib}
\usepackage{marginnote}
\usepackage{enumerate}
\usepackage{hyperref}
\usepackage{setspace}
\usepackage{subfig}
\newcommand*{\myTagFormat}[2]{(\ref{#1}$#2$)}
\usepackage{color, colortbl}
\definecolor{LightCyan}{rgb}{.5,1,0}
\definecolor{Blu}{rgb}{0,0.2,1}
\definecolor{Green}{rgb}{0,0.6,0.1}
\newcolumntype{P}[1]{>{\centering\arraybackslash}p{#1}}

\begin{document}

\title{Aircraft Loading Optimization: MemComputing the $5^{th}$ Airbus Problem}

\author{Fabio L. Traversa}
\email{email: ftraversa@memcpu.com}
\affiliation{MemComputing, Inc., San Diego, CA, 92130 CA}

%\author{Massimiliano Di Ventra}
%\email{email: diventra@physics.ucsd.edu}
%\affiliation{Department of Physics, University of California, San Diego, La Jolla, CA 92093}

%\collaboration{MUSO Collaboration}%\noaffiliation

%\author{Charlie Author}
% \homepage{http://www.Second.institution.edu/~Charlie.Author}
%\affiliation{
% Second institution and/or address\\
% This line break forced% with \\
%}%
%\affiliation{
% Third institution, the second for Charlie Author
%}%
%\author{Delta Author}
%\affiliation{%
% Authors' institution and/or address\\
% This line break forced with \textbackslash\textbackslash
%}%

%\collaboration{CLEO Collaboration}%\noaffiliation

\date{\today}% It is always \today, today,
             %  but any date may be explicitly specified

\begin{abstract}
On the January 22nd 2019, Airbus launched a quantum computing challenge to solve a set of problems relevant for the aircraft life cycle (\href{https://www.airbus.com/newsroom/press-releases/en/2019/01/airbus-launches-quantum-computing-challenge-to-transform-the-aircraft-lifecycle.html}{Airbus challenge web-page}). The challenge consists of a set of 5 problems that ranges from design to deployment of aircraft. This work addresses the $5^{th}$ problem. The formulation exploits an Integer programming framework with a linear objective function and the solution relies on the MemComputing paradigm. It is discussed how to use MemCPU$^{\text{TM}}$ software to solve efficiently the proposed problem and assess scaling properties, which turns out to be polynomial for meaningful solutions of the problem at hand. Also discussed are possible formulations of the problem utilizing non-linear objective functions, allowing for different optimization schemes implementable in modified MemCPU software, potentially useful for field operation purposes.             
\end{abstract}    

%\pacs{Valid PACS appear here}% PACS, the Physics and Astronomy
                             % Classification Scheme.
%\keywords{Suggested keywords}%Use showkeys class option if keyword
                              %display desired
                          
\maketitle

\section{Introduction}\label{Intro}

Automated computation in the 21st century has reached a paramount role in industry, finance, consumer technology and much more~\cite{Eiben2015,Kephart2003}. However, the more that computation becomes relevant, the greater the challenges presented to both industry and academia. Solutions to today's computational limits are provided by more and more sophisticated and high-performing CPUs, GPUs etc.~\cite{Vestias2014}, as well as by the insurgence of distributed computing in cloud infrastructures~\cite{Hashem2015}.

Nevertheless, the unceasing growth of computing demand is pushing towards completely new architectures as well as new paradigms. This can mean not only ``handling more computation,'' but also making the computation more efficient. For example, neuromorphic computing based on diverse realizations of artificial neural networks~\cite{Burr2016,Torrejon2017,Furber2014} promises a paradigm shift in machine learning and artificial intelligence.     

In this scenario, new computing architectures based on quantum physics and not strictly on algorithmic approaches promise to solve among the most challenging problems ranging from drug discovery to A~I\cite{Linke2017,Bennett2000,Ladd2010,Raha2007}. However, it is still not clear if and when reliable quantum computers will show what currently goes under the name of {\it quantum supremacy} over {\it classical} computing architectures~\cite{Harrow2017,MITtechrev,Dyakonov2019,Tang2018}.    

However, companies and academia are already looking at applications for future quantum computers~\cite{Accenture_report,MITtechrev,AribusURL}. I focus here on the challenge recently launched by Airbus~\cite{AribusURL} consisting of 5 problems related to the life cycle of an aircraft, i.e., ranging from the design to the deployment of an aircraft. The general requirements for the challenge are that for each problem a suitable algorithm implementable on a quantum computer should be developed, tested on a simulated quantum computer, and finally assessed in terms of the performance at scale. 

I consider in this work the 5th Airbus problem~\cite{AribusURL}: the Aircraft Loading Optimization (ALO) problem. The ALO requires optimizing the placement of containers of different sizes and weights in an aircraft subject to limitations on maximum weight allowed, maximum tolerable shear and center of gravity. However, instead of developing on algorithm for future quantum computers, I present a formalization of the problem using Integer Programming (IP) framework~\cite{Schrijver1998}. IP is a problem formulation widely used in industry and academia consisting of an objective function to be minimized over a set of variables. Variables may be constrained to take binary, integer or continuous values. Moreover, the objective function may be subject to further constraints in the form of linear inequalities among variables. Finally, depending on the nature of the objective function, we can have different IP problems such as linear, quadratic or other non-linear type.

The formulation I consider here covers all requirements of the ALO statement and leads to an IP problem involving binary variables only and having two objective functions to be minimized, one linear and the other non-linear. I therefore propose two different solution strategies to solve this multi-objective IP problem.

The binary IP (BP) problem is NP-complete, sometimes also known as {\it 0-1 linear programming}, and is one of Karp's 21 NP-complete problems~\cite{complexity_bible}. Therefore, it is not surprising that Airbus poses a challenge about a problem whose natural formulation leads to a special case of on NP-complete problem. Several general-purpose open source~\cite{GLPK,Forrest2018,Gleixner2018,Ralphs2017,Berkelaar2004} and commercial solvers~\cite{Gurobi,CPLEX,FICO,MATLAB} have been developed to solve IP problems. However, these can fail when the problems are particularly hard and therefore specialized solvers optimized to exploit specific structures for ILP have also been developed~\cite{applegate2006concorde,Floudas2005,Boston99,sorensen2014hybridizing}.

\begin{figure*}[t!]
	\centerline{\includegraphics[width=1.6\columnwidth]{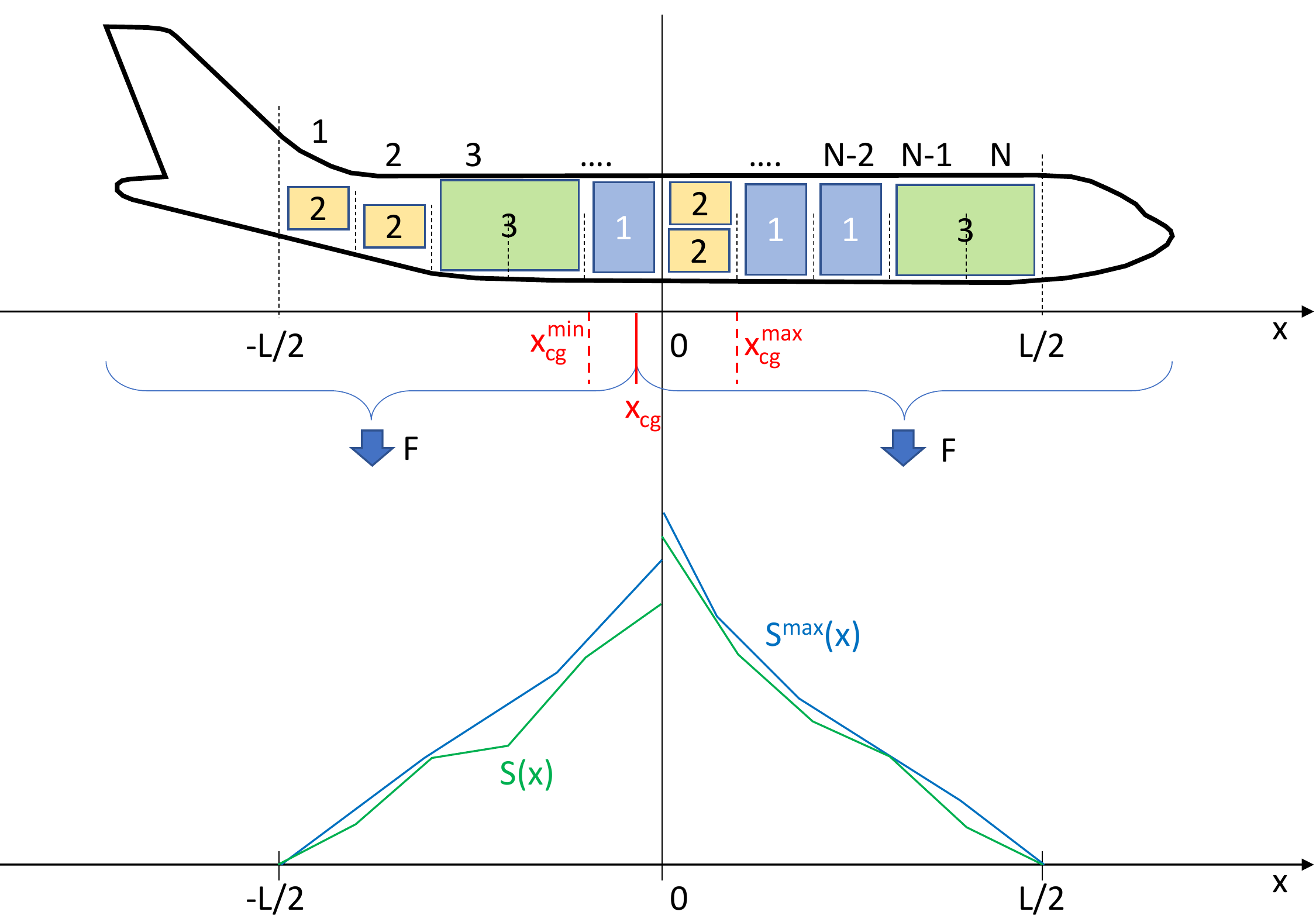}}
	\caption{Sketch of the Airbus Aircraft Loading Optimization problem. The aircraft is to be loaded by selecting a portion of the available payload. The containers can have different shapes. For this ALO, 3 different shapes are considered numbered from 1 to 3 as in the figure. The maximum weight allowed is limited. The center of gravity $x_{cg}$ of the system aircraft + containers must remain within the limit $[x^{\min}_{cg}, x^{\max}_{cg}]$ and the shear curve $S(x)$ must also be limited to remain under a give shear limit curve $S^{\max}(x)$.}
	\label{figALO}
\end{figure*}

IP problems can be approached employing different techniques. A classification of those includes heuristics~\cite{Danna2004,Boston99,Glover1977,Traversa2018a} and exhaustive algorithms~\cite{Schrijver1998,barnhart1998branch,benders1962partitioning,hooker2003logic} also called ``complete algorithms.'' The complete algorithm for ILP that is most commonly employed is a 
combination of cutting planes and branch-and-bound, also known as the branch-and-cut algorithm~\cite{Schrijver1998}. All these solvers have demonstrated varied degrees of success on a variety of IP problems~\cite{abara1989applying,kroon2009new,stahlbock2008operations,melo2009facility,bard2003staff}, however scaling properties for BP problems such as the one proposed by Airbus can easily show exponential blowing-up when applying one of those methods~\cite{complexity_bible}.

In this work I discuss the general purpose approach to the solution of the Aibus ALO problem based on the {\it memcomputing paradigm} introduced in \cite{UMM}. Memcomputing (computing {\it with} and {\it in} memory~\cite{Di_Ventra2018}) approach is neither based on stochastic search nor on trial and error strategy. In addition, memcomputing does not use a set of instructions recursively employed to find solutions to the problem at hand. Therefore, we can classify the memcomputing approach as non-algorithmic~\cite{Di_Ventra2018}.     

The memcomputing approach relies on embedding a given problem into an electronic circuit, which represents a possible realization of a memcomputing machine (MM)~\cite{UMM,DMM2,Di_Ventra2018}. These electronic circuits, when designed in order to satisfy crucial properties~\cite{DMM2}, naturally relax to an equilibrium that represents the solution of the problems at hand~\cite{DMM2,Di_Ventra2018,no-chaos,noperiod}. 

In order to approach the Airbus ALO problem I use the MM realization described in \cite{Traversa2018a} employing {\it self-organizing algebraic gates} (SOAGs). The SOAG-based MM can be efficiently simulated in software and in this work I use the software developed at MemComputing Inc.~\cite{MemWeb} named MemCPU coprocessor. 

In section~\ref{ALO_statement_section} I briefly describe the Airbus ALO problem. In section~\ref{ALO_BP_section} the BP formulation of the ALO problems is discussed. In section~\ref{Mem_section} a quick overview of memcomputing and MemCPU software is provided, while in section~\ref{results_section} are discussed both numerical results and scaling properties of MemCPU software applied to the ALO problem. 

\section{Airbus Aircraft Loading Optimization Problem}\label{ALO_statement_section}

The ALO problem proposed by Airbus is sketched in Fig.~\ref{figALO}. Following the problem statement we need to optimize the placement of a subset of containers picked from an available payload formed by $n$ different containers. Each container is described by the triplet $(k,m_k,s_k)$ with $k$ the identification number, $m_k$ the mass and $s_k$ the size of the $k-th$ container. 

The aircraft has $N$ available positions for standard cargo as depicted in Fig.~\ref{figALO}. The size can be of 3 different types: size 1 occupies one position; size 2 occupies half a position and may share the position with another size 2 container; size 3 occupies 2 positions.    

The problem requires maximizing the mass of the carried freight for the flight such that:
\begin{enumerate}[(A)] 
\item the containers must be placed consistently with size and positions available 
\item the mass of the carried freight does not exceed the maximum payload capacity of the aircraft $W_p$
\item the center of gravity position of the carried freight + aircraft must be within the limits $[x^{\min}_{cg}, x^{\max}_{cg}]$
\item the shear curve $S(x)$ defined by the container distribution must be bounded by a given maximal shear curve $S^{\max}(x)$
\item while maximizing the mass, the position of the center of gravity of the carried freight + aircraft should be optimized to be as close as possible to a target center of gravity $x^t_{cg}$. 
\end{enumerate}

These constraints are sketched in Fig.~\ref{figALO}. While the constraints (A) and (B) do not need additional details or explanations, the others do. In (C) the center of gravity is defined by the the equilibrium of all forces (container + aircraft) in the gravity axis direction (see force $F$ in Fig.~\ref{figALO}). Each container is supposed to have uniform mass, therefore the weight force can be considered as the weight of a point located at the center of the container. Finally, the mass and location of the center of gravity of the empty aircraft are provided.

The constraint (D) requires a consistent definition of the shear curve. The shear curve is defined by
\refstepcounter{equation}\label{ShearCurve}
\begin{align}
   S(x)=&\int^x_{-\frac{L}{2}}m(x')dx'\text{ }\text{  for  }\text{ }x<0\tag*{\myTagFormat{ShearCurve}{a}}\\
   S(x)=&\int^{\frac{L}{2}}_xm(x')dx'\text{ }\text{  for  }\text{ }x>0\tag*{\myTagFormat{ShearCurve}{b}}
\end{align}
where $L$ is the length of the loading region of the aircraft and $x$ is the position relative to the center of the loading zone (see Fig.~\ref{figALO}). In the statement of the problem, the maximum shear curve $S^{\max}(x)$ can be either linear and symmetric with respect to $x=0$, or asymmetric or some non-linear function of $x$.

Finally, (E) represents an extra optimization requirement and not really a constraint. In fact, from the Airbus statement, what should be primarily optimized is the mass of the carried freight, then optimize the distribution such that the resulting center of gravity is as close as possible to a target.

\section{Integer Programming Formulation}\label{ALO_BP_section}

A natural way to mathematically formulate the problem posed in section~\ref{ALO_statement_section} is using Integer Programming framework~\cite{Schrijver1998}. 

Let us start assigning a binary variable $y_{k,j}$ to each container $k=1,...,n$ located in the position $j=1,...,N$ of the aircraft. Since each container can be located in at most one position in the aircraft we constrain these variables requiring
\refstepcounter{equation}\label{constraints}
\begin{equation}
\sum_{j=1}^N y_{k,j}\leq 1\text{ }\text{ for each } k=1,...,n\tag*{\myTagFormat{constraints}{a}}.
\end{equation}

On the other hand, we have the constraint on the size and the number of available spots on the aircraft that can be expressed as a non-linear inequality. In order to formalize this constraint, let us further characterize the variable $y_{k,j}$. We consider here, for simplicity, only the 3 sizes defined in the Airbus statement. However, this formulation can be easily extended to more complicated sizes and bin distributions in the aircraft. Let us define $K_1, K_2$ and $K_3$ as three sets of indexes such that $K_1\cup K_2\cup K_3 = \{1,...,n\}$. Moreover, if $k\in K_h$ then $s_k=h$. Therefore, $K_1, K_2$ and $K_3$ regroup the variables in sets of variables corresponding to the same container size.     

Using this index characterization, the constraint (A) on the sizes and bins can be expressed as
\begin{align}
\sum_{k\in K_1} y_{k,1} + \frac{1}{2}\sum_{k\in K_2} y_{k,1} + \sum_{k\in K_3} y_{k,1} \leq 1\tag*{\myTagFormat{constraints}{b'}}\\
\sum_{k\in K_1} y_{k,j} + \frac{1}{2}\sum_{k\in K_2} y_{k,j} + \sum_{k\in K_3} (y_{k,j-1}+y_{k,j}) \leq 1\nonumber\\
\text{ }\text{  for  }\text{ }j=2,...,N-1\tag*{\myTagFormat{constraints}{b''}}\\
\sum_{k\in K_1} y_{k,N} + \frac{1}{2}\sum_{k\in K_2} y_{k,N} + \sum_{k\in K_3} y_{k,N-1} \leq 1\tag*{\myTagFormat{constraints}{b'''}}
\end{align}

It is easy to prove that these inequalities guarantee that for each bin there is no overlapping of containers except for containers of size 2 for which two of them can occupy the same bin (the coefficient $1/2$ allows the overlapping). The case of size 3 (containers that occupy two consecutive bins) is enforced by the term $\sum_{k\in K_3} (y_{k,j-1}+y_{k,j})$ in which two consecutive containers of size 3 appear in the same constraint; therefore a container of size 3 occupies 2 bins if selected. It is also worth noticing that containers of size 3 have $j$ that ranges only from 1 to $N-1$ because, since they occupy two bins, there ore only $N-1$ possible locations for them.

The constraint (B) can be trivially described by
\begin{equation}
\sum_{k,j}m_ky_{k,j}\leq W_p\tag*{\myTagFormat{constraints}{c}}.
\end{equation}

The constraint (C) on the center of gravity requires the definition of signed distance $d_{s_k,j}$ from $x=0$ for a container of size $s_k$ located in the bin $j$. Considering equally distributed bins around $x=0$ (this assumption is the same reported in the Airbus statement but can be easily relaxed), we have that 
\refstepcounter{equation}\label{length}
\begin{align}
d_{s_k,j} =& \dfrac{2j-N-1}{2N}L&\text{ }\text{  for  }\text{ }s_k &= 1,2 \tag*{\myTagFormat{length}{a}}\\
d_{s_k,j} =& \dfrac{2j-N}{2N}L&\text{ }\text{  for  }\text{ }s_k &= 3 \tag*{\myTagFormat{length}{b}}
\end{align}

Using this distance definition, the center of gravity of the loaded aircraft can be evaluated as 
\begin{equation}
x_{cg} = \dfrac{\sum_{k,j}m_kd_{s_k,j}y_{k,j}+W_ex^e_{cg}}{\sum_{k,j}m_ky_{k,j}+W_e}\label{xcg}
\end{equation}
where $W_e$ and $x^e_{cg}$ are respectively the mass and the center of gravity of the empty aircraft.      

Eq.~\eqref{xcg} allows the formalization of the constraint (C) through the inequalities
\begin{align}
\sum_{k,j}m_k(d_{s_k,j}-x^{\max}_{cg})y_{k,j}\leq W_e(x^{\max}_{cg}-x^e_{cg})\tag*{\myTagFormat{constraints}{d'}}\\
\sum_{k,j}m_k(x^{\min}_{cg}-d_{s_k,j})y_{k,j}\leq W_e(x^e_{cg}-x^{\min}_{cg})\tag*{\myTagFormat{constraints}{d''}}
\end{align}
that express $x_{cg}\leq x^{\max}_{cg}$ and $x_{cg}\geq x^{\min}_{cg}$ respectively.

Regarding the constraint (D), the shear curve can be easily calculated at each bin location and set smaller than $S^{\max}(x)$:   
\begin{align}
\sum_{k\in K_1\cup K_2} \sum_{j'=1}^j m_ky_{k,j'}& +\nonumber\\
+\sum_{k\in K_3} &\left(\sum_{j'=1}^{j-1}m_ky_{k,j'} +\frac{1}{2}m_ky_{k,j}\right)\leq\nonumber\\ \leq &S^{\max}(x_j)\text{ }\text{  for  }\text{ }j\leq N/2   \tag*{\myTagFormat{constraints}{e'}}\\
\sum_{k\in K_1\cup K_2} \sum_{j'=j}^{N} m_ky_{k,j'}& +\nonumber\\
+\sum_{k\in K_3} &\left(\sum_{j'=j-1}^{N-1}m_ky_{k,j'} +\frac{1}{2}m_ky_{k,j}\right)\leq\nonumber\\ \leq &S^{\max}(x_j)\text{ }\text{  for  }\text{ }j\geq N/2   \tag*{\myTagFormat{constraints}{e''}}
\end{align}
which expresses the shear curve defined in Eq.~\eqref{ShearCurve} at the centers of the bins $x_j$ taking into account the different sizes of the containers. 

Finally we can define the objective function of the ALO problem as
\begin{equation}
f(y) = -\sum_{k,j}m_ky_{k,j}\label{objective}.
\end{equation}
whose minimization over $y$, subject to the collection of constraints \eqref{constraints}, provides the distribution and the maximum mass of the carried freight satisfying the constraints (A)-(D) of section~\ref{ALO_statement_section}.

\subsection{Center of Gravity Optimization}\label{CG_optimization_section}

We briefly discuss in this section the further optimization required in (E) of section~\ref{ALO_statement_section}.
This can be handled with the following scheme. Once the problem \eqref{objective} is solved, we obtain the maximum mass of the carried freight as 
\begin{equation}
W^{\max} = \sum_{k,j}m_k\tilde y_{k,j}\label{maxmass}
\end{equation}
where $\tilde y$ is the solution of the minimization of \eqref{objective} subject to \eqref{constraints}.
We therefore define a new IP problem where we include the constraints (\ref{constraints}a,\ref{constraints}b,\ref{constraints}c,\ref{constraints}e) but exclude for now (\ref{constraints}d). Instead we include the extra constraint 
\begin{equation}
-\sum_{k,j}m_ky_{k,j}\leq -\tau W^{\max}\label{extraconstraint}.
\end{equation}
where $0\leq \tau\leq$ is a tolerance parameter that can be used in case we are interested in a slightly lighter carried freight but better center of gravity. 

We then define the new objective function
\begin{equation}
f(y) = \left\{ {\begin{array}{cc}
	x_{cg}- x^t_{cg} & \text{ if } x_{cg}\geq x^t_{cg}\\
	-x_{cg}+ x^t_{cg} &  \text{ if } x_{cg}\leq x^t_{cg} \\
	\end{array} } \right.\label{newobjective}
\end{equation}
with $x_{cg}$ defined by Eq.~\eqref{xcg}. 

The target now is to find the assignment of $y$ that minimizes Eq.~\eqref{newobjective} under the constraints (\ref{constraints}a,\ref{constraints}b,\ref{constraints}c,\ref{constraints}e) and \eqref{extraconstraint}. This can be achieved either using a sequence of linear IP problems in which we reintroduce the constraint (\ref{constraints}d) and we iteratively shrink the $[x^{\min}_{cg}, x^{\max}_{cg}]$ around $x^t_{cg}$, or by implementing it directly as a non-linear IP problem with objective function Eq.~\eqref{newobjective}.

\section{MemComputing Approach}\label{Mem_section}

The memcomputing approach to IP problems is based on the concept of Self-Organizing Algebraic Gates (SOAGs) introduced in~\cite{Traversa2018a}. SOAG is a novel circuit design developed at MemComputing, Inc.~\cite{MemWeb} by the author of this work and is part of a class of self-organizing circuits as Self-Organizing Logic Gates (SOLGs) introduced in~\cite{DMM2,patentSOLC}. 

Both SOLGs and SOAGs are building blocks for practical realizations of Universal Memcomputing Machines (UMM)~\cite{Di_Ventra2018,UMM,traversaNP}, with digital input-output (Digital MM, DMM)~\cite{DMM2}.

\begin{figure}%[t!]
	\centerline{\includegraphics[width=1.02\columnwidth]{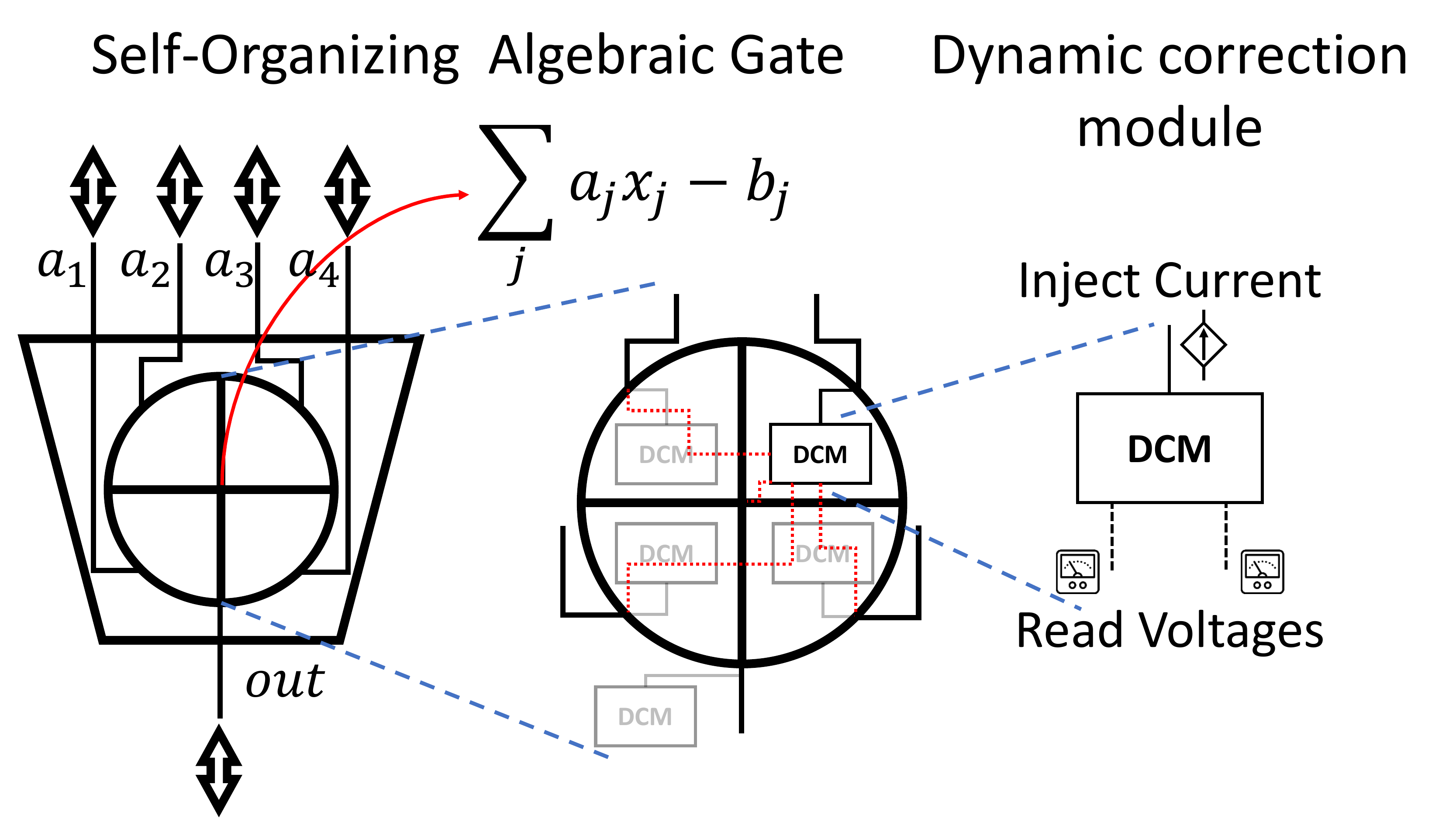}}
	\caption{Reprint from~\cite{Traversa2018a}. Sketch of a Self-Organizing Algebraic Gate. All terminals allow a superposition of incoming and outgoing signals from the surrounding circuit. The central unit processes the signals in order to satisfy a linear algebraic relation consistent with the requirement of the ``out'' terminal. The self-organization is enforced by the Dynamic Correction Modules that read voltages from all terminals and inject a current to the appropriate terminal as long as the algebraic relation is not satisfied.}
	\label{figSOAG}
\end{figure}
\begin{figure}[t!]
	\centerline{\includegraphics[width=1.02\columnwidth]{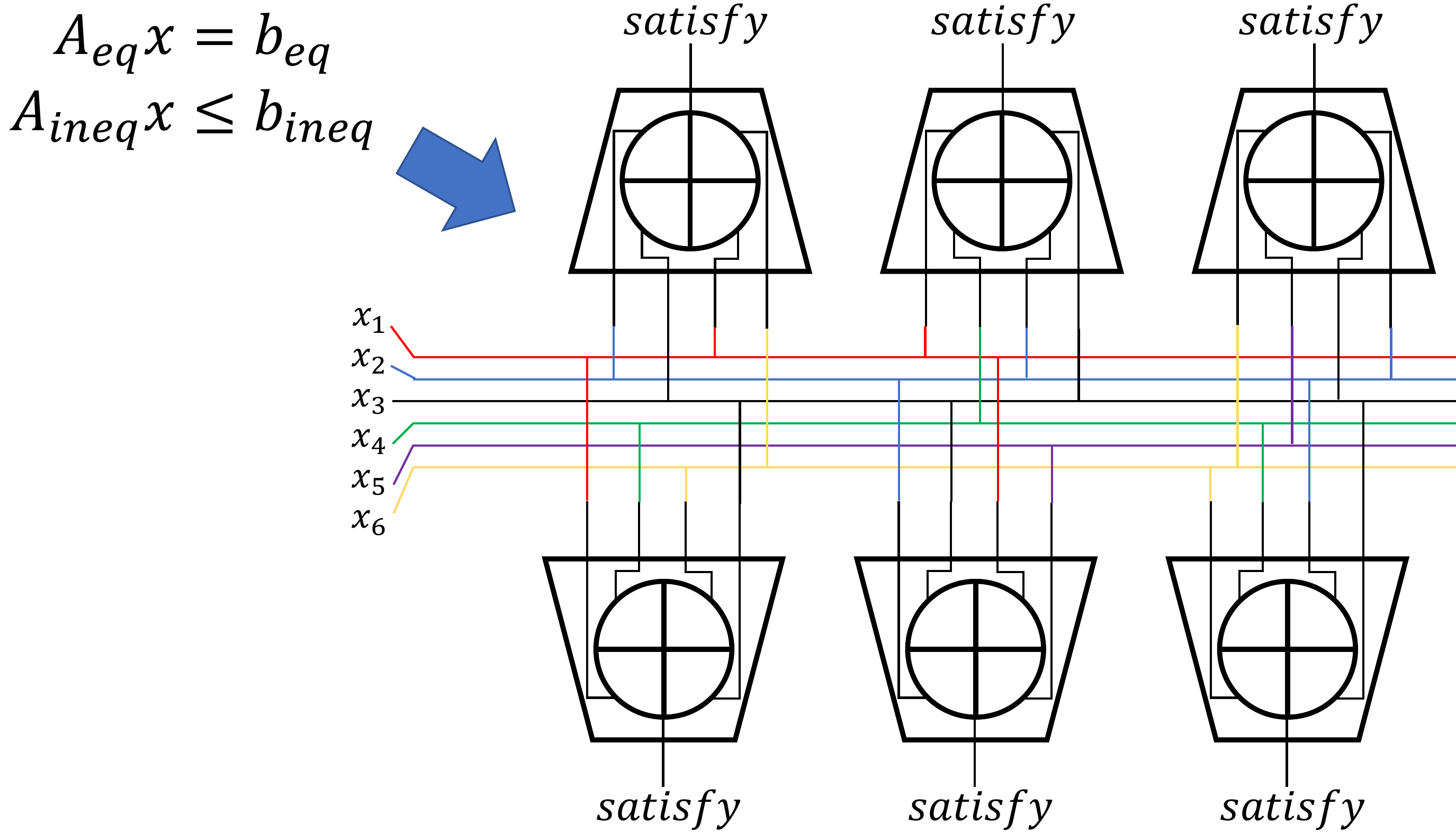}}
	\caption{Reprint from~\cite{Traversa2018a}. Sketch of a Self-Organizing Algebraic Circuit (SOAC). SOAGs are connected together in an architecture that directly maps the IP into the SOAC.\label{figSOAC}}
\end{figure}

The SOAG is designed to self-organize toward an {\it algebraic relation} representing a linear inequality among variables. In this work, the SOAGs are designed to satisfy linear relations between binary variables (see Fig.~\ref{figSOAG}) since the ALO problem formulation involves binary variables only.     

By connecting together SOAGs we form a Self-Organizing Algebraic Circuit (SOAC), see Fig.~\ref{figSOAC}. The SOAC collectively self-organizes in order to satisfy the relations embedded in the gates. In this way, it is trivial to embed an IP problem directly into the SOAC. Each inequality in \eqref{constraints} or \eqref{extraconstraint} can be mapped directly into a SOAG. The objective functions \eqref{objective}  can be easily reformulated as an extra linear inequality 
\begin{equation}
\sum_{k,j}f_{k,j}y_{k,j}\le \tilde b\label{objectiveremapped}
\end{equation}
where $\tilde b$ represents a threshold for the maximum carried freight that is dynamically changed each time a feasible solution is found in order to find solutions closer and closer to the global minimum of the problem.  
On the other hand, the objective function \eqref{newobjective} can be implemented as a pair of linear inequalities of the form
\begin{align}
\sum_jf^+_{k,j}(\tilde b)y_{k,j}\le g^+(\tilde b)\label{objectiveremapped1}\\
\sum_jf^-_{k,j}(\tilde b)y_{k,j}\le g^-(\tilde b)\label{objectiveremapped2}
\end{align}
where $\tilde b$ is an extra parameter that again can be dynamically changed to find solutions of increasing quality, each time closer to the global optimum. $f^{\pm}$ and $g^{\pm}$ are linear functions of $\tilde b$ defined by substituting \eqref{xcg} in \eqref{newobjective}. This also leads to defining $\tilde b$ as the threshold of the distance of the actual center of gravity from the target $x^t_{cg}$. 

The problem formulated and embedded in a SOAC as described can be efficiently handled by either actually building the electronic circuit or just by simulating it since it involves only standard (non-quantum) electronic components. Simulating means solving differential equations of the form 
\begin{equation}
\dfrac{dQ(i,v,x)}{dt} = F(i,v,x),\label{DAE}
\end{equation}
with appropriate initial conditions and where $Q$ and $F$ are non-linear functions of the voltages ($v$), currents ($i$) and extra internal state variables ($x$) characterizing the circuit. In this picture a configuration of the voltages ($v$) at a given time represents an actual assignment to the variables of the IP problem by reading the voltages through thresholds: if a voltage at a node of the circuit is above the threshold, then it corresponds to a logical 1, and otherwise corresponds to a logical 0. On the other hand, the transition function of these machines (namely the function that maps input to output) is physical (analog) and takes full advantage of the {\it collective state} of the system to process information~\cite{collective,traversaNP,DMM2}. However, despite its analog nature, the mapping of voltages into binary variables through thresholds allows efficient size scaling for these machines, avoiding the bottleneck related to the precision of writing and reading inputs and outputs. Finally, it is worth noticing here that the memcomputing approach used to solve binary problems as presented in~\cite{Traversa2018a} for the case of IP problems, does not provide proof of optimality for a given solution, nor does it detect the infeasibility of a problem.    

In this work we consider the simulation of the SOAC implemented in the MemCPU software developed at MemComputing, Inc.\cite{MemWeb} and available also as a Software as a Service offering.  

The working principle of SOAGs and SOACs, i.e., self-organization of voltages and currents of an electronic circuit in order to satisfy algebraic relations, is enforced by both active and passive electronic elements with and without memory~\cite{Traversa2018a,DMM2,Di_Ventra2018}. The features of DMMs have been investigated, and interesting properties emerge from a correct design such as long-range order correlations and topological robustness~\cite{topo,Bearden2018}. If mathematical requirements described in~\cite{DMM2} are fulfilled, then persistent oscillations or chaos are avoided~\cite{no-chaos,noperiod}. Self-organizing digital circuits (i.e. both SOLCs and SOACs), being a realization of DMMs can in principle solve efficiently (i.e. with polynomial resources) complex combinatorial optimization problems, given that mathematical features on the constitutive equation~\eqref{DAE} are satisfied~\cite{UMM,DMM2,Di_Ventra2018}. Self-organizing digital circuits (i.e. both SOLCs and SOACs) have also been proved to efficiently handle a variety of combinatorial optimization problems ranging from maximum satisfiability (MAXSAT)~\cite{Traversa2018,Sheldon2018} to quadratic unconstrained binary optimization (QUBO)~\cite{spinglass} and from IP~\cite{Traversa2018a} to the training of neural networks~\cite{AcceleratingDL}. 

\section{Scaling Results}\label{results_section}

In order to assess the scaling properties of solving the Airbus ALO problem employing MemCPU coprocessor we need to generate a set of meaningful benchmarks at different values of $N$ and $n$. Since we have no information about the actual values of $N$ and $n$ and not even on the ratio between number of containers at different sizes or typical mass distribution of the containers available, we use the sample data set provided by Airbus (see Table~\ref{Airbusdataset} for the data set, or it can also be found from~\cite{AribusURL}) to extrapolate. Notice that the data set provided from Airbus is just ``for illustration and for testing the algorithm''~\cite{AribusURL}. 

\begin{table}
	\begin{center}
		\begin{tabular}{ |P{1.5cm}|P{1.5cm}|P{3.5cm}|  }	
			\hline \hline
\multicolumn{3}{|p{6.5cm}|}{$N=20$} \\
		%	\hline
\multicolumn{3}{|p{6.5cm}|}{$n=30$} \\
		%	\hline
\multicolumn{3}{|p{6.5cm}|}{$W_p=40000$} \\
		%	\hline
\multicolumn{3}{|p{6.5cm}|}{$W_e=120000$} \\
		%	\hline
\multicolumn{3}{|p{6.5cm}|}{$x^e_{cg}=-0.05L$} \\
		%	\hline
\multicolumn{3}{|p{6.5cm}|}{$x^{\min}_{cg}=-0.1L$} \\
		%	\hline
\multicolumn{3}{|p{6.5cm}|}{$x^{\max}_{cg}=0.2L$} \\
		%	\hline
\multicolumn{3}{|p{6.5cm}|}{$x^t_{cg}=0.1L$} \\
		%	\hline
\multicolumn{3}{|p{6.5cm}|}{$S^{\max}(x=0)=22000$, linear, symmetric} \\ 
			\hline \hline
$k$ 	&	$s_k$ 	&	$m_k$ (kg) \\
\hline\hline
1	&	 1 	&	2134 \\
\hline 
2 	&	1 	&	3455  \\
\hline
3 	&	1 	&	1866  \\
\hline
4 	&	1 	&	1699 \\
\hline
5 	&	1 	&	3500 \\
\hline
6 	&	1 	&	3332  \\
\hline
7 	&	1 	&	2578  \\
\hline
8 	&	1 	&	2315  \\
\hline
9 	&	1 	&	1888  \\
\hline
10 	&	1	&	1786  \\
\hline
11 	&	1 	&	3277  \\
\hline
12 	&	1 	&	2987  \\
\hline
13 	&	1 	&	2534  \\
\hline
14 	&	1 	&	2111 \\
\hline
15 	&	1 	&	2607  \\
\hline
16 	&	1 	&	1566 \\
\hline
17 	&	1 	&	1765 \\
\hline
18 	&	1 	&	1946 \\
\hline
19 	&	1 	&	1732 \\
\hline
20 	&	1 	&	1641 \\
\hline
21 	&	2 	&	1800 \\
\hline
22 	&	2 	&	986   \\
\hline
23 	&	2 	&	873 \\
\hline
24 	&	2 	&	1764 \\
\hline
25 	&	2 	&	1239 \\
\hline
26 	&	2 	&	1487 \\
\hline
27 	&	2 	&	769 \\
\hline
28 	&	2 	&	836 \\
\hline
29 	&	2 	&	659 \\
\hline
30 	&	2 	&	765 \\
\hline
		\end{tabular}
\end{center}
\caption{Airbus ALO data set~\cite{AribusURL}.  \label{Airbusdataset}}
\end{table}

\begin{figure}[t!]
	\includegraphics[width=\columnwidth]{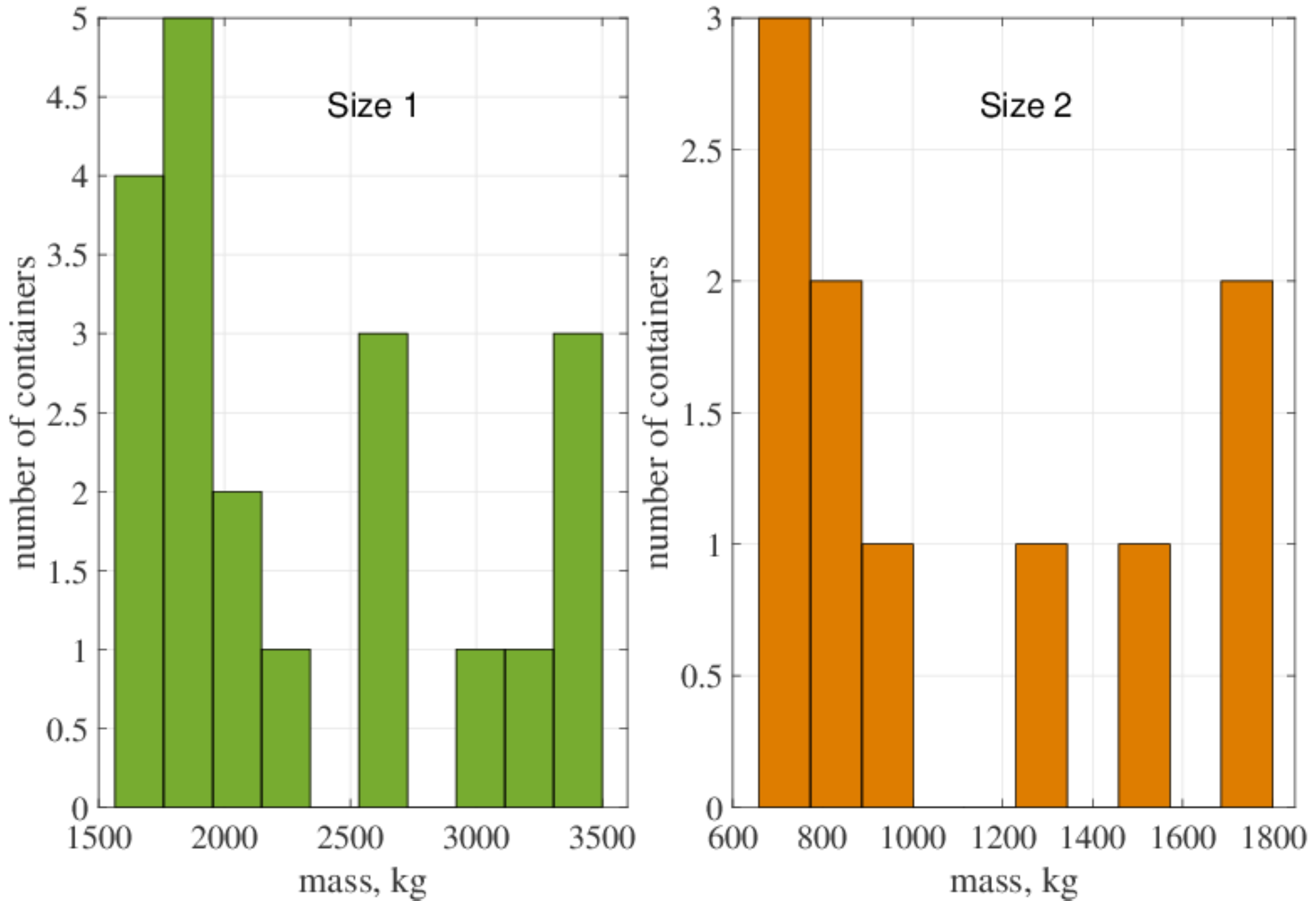}
	\caption{Histograms of container mass from Airbus data set reported in Table~\ref{Airbusdataset}. \label{distribution_data_set}}
\end{figure}
In order to generate benchmarks for different sizes we need to generate a distribution of containers. One of the possible ways is to try to produce a distribution similar to that from the sample data set. In Fig.~\ref{distribution_data_set} the distributions of container masses from the data set in Table~\ref{Airbusdataset} is reported. Even though there are not enough data to determine with certainty the distribution shape, it is reasonable to assume that both distributions can be recreated from a bimodal Gaussian distribution, cut off at certain boundaries.       

\begin{figure*}[t!]
	\centerline{\includegraphics[width=2.1\columnwidth]{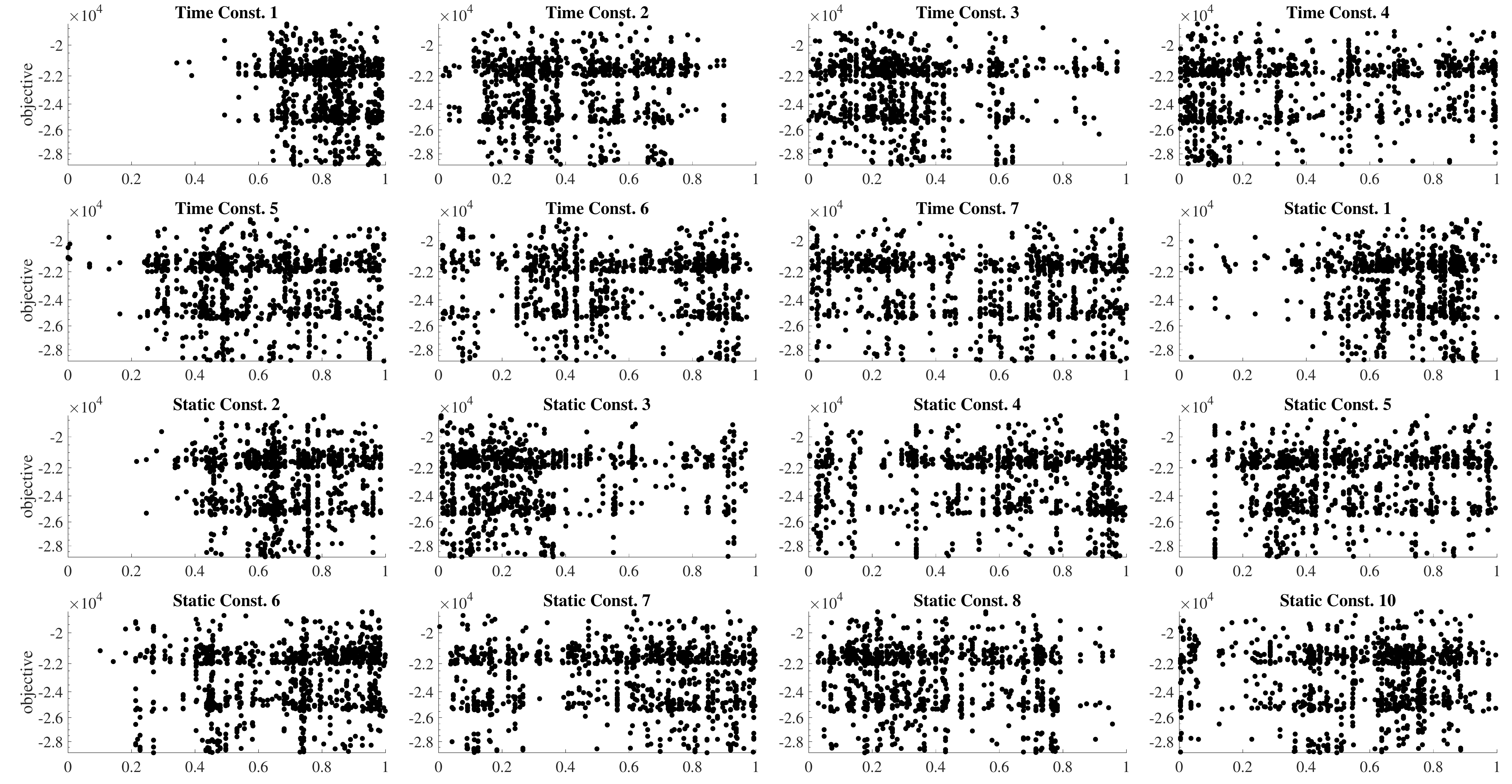}}
	\caption{Monte Carlo distribution of the objective \eqref{objective} versus MemCPU coprocessor parameters~\cite{MemWeb}. The ALO problem is defined with the parameters in Table~\ref{Airbusdataset} and $x^{\min}_{cg}=-0.006$. Number of Monte Carlo iterations = 1200, simulation time = 8 and number of Markov Chains = 8. \label{MCdistribution}}
\end{figure*}

In Fig.~\ref{matlabcode} I have included Matlab code that generates ALO problems for any $n$ and $N$ (it is also available a converter from Matlab to .mps format at \cite{LuongMatlab}). The distribution of containers is generated in the subfunction {\fontfamily{qcr}\selectfont container\_distribution\_generator}. A bimodal Gaussian distribution of containers is generated for each size type. The ranges are chosen in such a way that the distributions in Fig.~\ref{distribution_data_set} can be qualitatively reproduced. It is also worth noticing that the masses are scaled depending on $N$. In fact, maintaining the same weight of the airplane and the same limit on the maximum weight, but increasing the number of bins, requires for consistency that each bin be scaled in size, and therefore we scale all container masses accordingly.       

\subsection{MemCPU software results}\label{MemCPU_section}

\begin{table}
	\begin{center}
		\begin{tabular}{ |p{2.5cm}|P{1.4cm}||p{2.6cm}|P{1.4cm}|  }	
			\hline \hline
			\multicolumn{2}{|P{3.9cm}||}{Time Constants} & \multicolumn{2}{|P{3.9cm}|}{Static Constants} \\
			\hline\hline
			Time Const. 1	&	 0.79293	&	 Static Const. 1	&	 0.86027\\
			\hline
			Time Const. 2	&	 0.28953	&	 Static Const. 2	&	 0.43655\\
			\hline
			Time Const. 3	&	 0.19784	&	 Static Const. 3	&	 0.24784\\
			\hline
			Time Const. 4	&	 0.85076	&	 Static Const. 4	&	 0.68352\\
			\hline
			Time Const. 5	&	 0.42802	&	 Static Const. 5	&	 0.37106\\
			\hline
			Time Const. 6	&	 0.39473	&	 Static Const. 6	&	 0.44440\\
			\hline
			Time Const. 7	&	 0.76838	&	 Static Const. 7	&	 0.26562\\
			\hline
			            	&	        	&	 Static Const. 8	&	 0.21645\\
			\hline
			            	&	        	&	 Static Const. 9	&	 1.00000\\
			\hline
			            	&	        	&	 Static Const. 10	&	 0.88580\\	
			\hline
		\end{tabular}
	\end{center}
	\caption{Parameters for MemCPU coprocessor~\cite{MemWeb} extracted from the distribution of Fig.~\ref{MCdistribution}.  \label{parameters}}
\end{table}
\begin{figure}[t!]
	\includegraphics[width=\columnwidth]{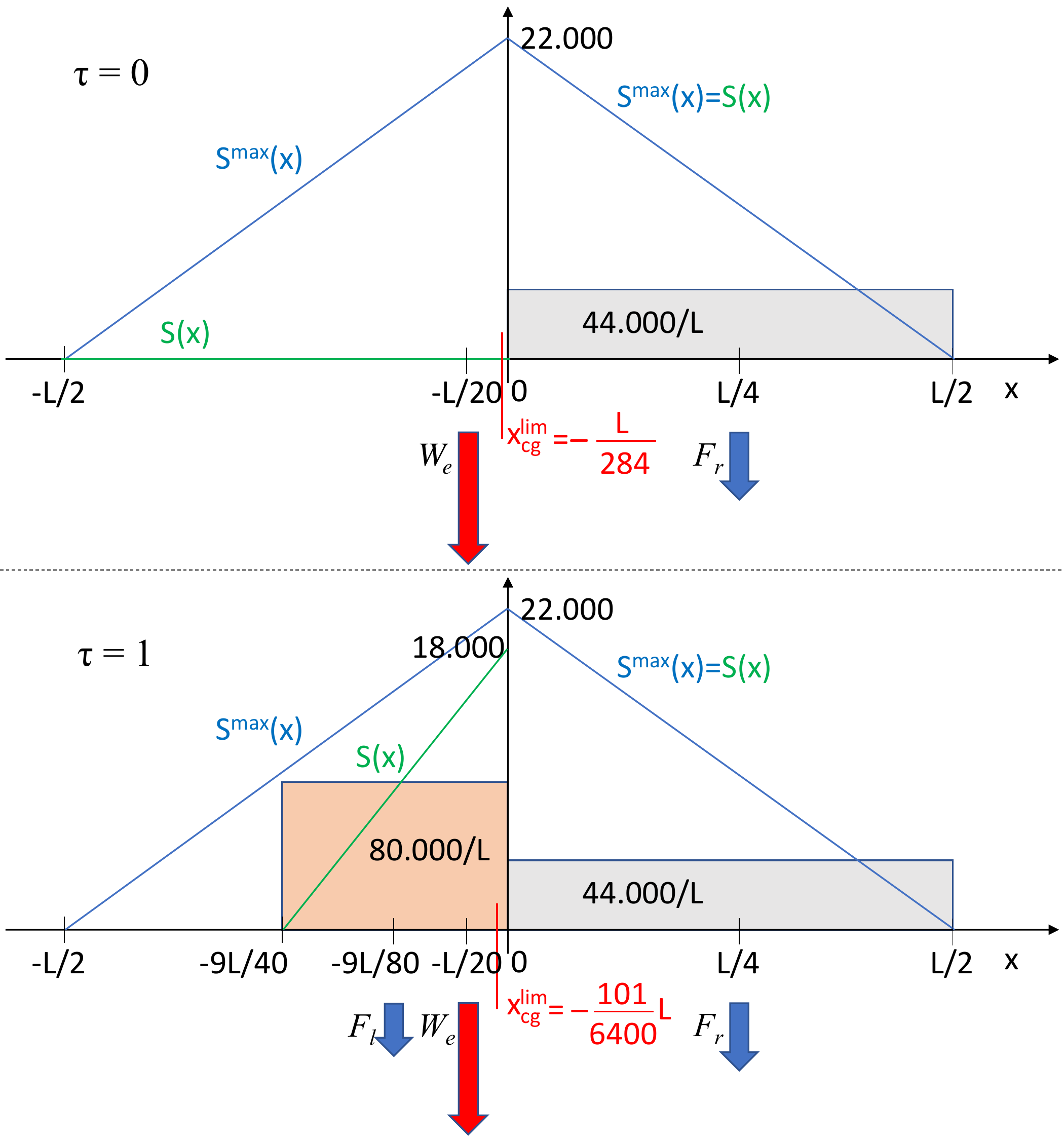}
	\caption{Configuration of forces to evaluate how close the center of gravity $x^{\lim}_{cg}$ can approach the target $x^t_{cg}=0.1$ (see Table~\ref{Airbusdataset}), without restriction on the minimum selected payload ($\tau=0$) and with maximum selected payload ($\tau=1$). For the maximum selected payload, the maximum density of $80000/L$ (container mass per unit of aircraft length) is assumed, consistent with the data set in Table~\ref{Airbusdataset} and the subfunction {\fontfamily{qcr}\selectfont container\_distribution\_generator} in Fig.~\ref{matlabcode}.\label{xlimitcg}}
\end{figure}

As discussed in section~\ref{Mem_section}, a MemCPU software is an emulator of an electronic circuit with a given IP problem embedded within. However, in order to be efficient in solving the problem at hand, the MemCPU coprocessor needs a proper set of parameters characterizing electronic elements such as resistors, capacitors, etc. These parameters do not depend on the size of the problem (they are scale free) but depend on the structure of the problem. The MemComputing SaaS currently provides a Monte Carlo routine that evaluates the distribution of the objective function of the IP problem as a function of the parameters~\cite{tuning}; however future releases will provide a predictor routine for parameters that will avoid the Monte Carlo-based tuning~\cite{tuning}. The objective function distribution is reported in Fig.~\ref{MCdistribution}. This distribution has been computed for the problem generated using the code in Fig.~\ref{matlabcode} using the container distribution in Table~\ref{Airbusdataset} as input and $x^{\min}_{cg}=-0.006$. The choice of $x^{\min}_{cg}$ is only for tuning purposes because this restricts the possible feasible solutions of the problem making it ``harder.'' In fact, it is easy to estimate the limit of the center of gravity $x^{\lim}_{cg}$ for any selection of containers ($\tau=0$) from the available payload. In Fig.~\ref{xlimitcg}, the forces in the center of gravity evaluation (Eq.~\ref{xcg}) are summarized for the limit of containers of equal weight corresponding to the maximum weight allowed by the maximum shear curve. The value is $x^{\lim}_{cg}=-1/284\approx -0.0035$ and I chose $x^{\min}_{cg}=-0.006$ that largely reduces the number of feasible solutions. This choice for $x^{\min}_{cg}$ leads to minimal feasible solutions that have objective far from maximum allowed carried freight $W_p$, however, used for tuning is a good choice because largely reduces the number of feasible solutions. This choice helps to produce a sharper distribution of objective function values versus parameters which also is useful for solving the problem in section~\ref{CG_optimization_section}.     

From the objective distribution of Fig.\ref{MCdistribution}, a set of the parameters for the MemCPU coprocessor can be easily extracted and used for running any other instance of the ALO problem. However, extracting a good set of parameters from the distribution can be done in multiple ways. In this work, the following procedure has been followed: from each Markov chain the set of parameters returning the best objective is selected. If there is more than one set of parameters with the same objective for the same Markov chain, the choice is made based upon the one that has the best average of the objectives related to the closest 10 sets of parameters within the same Markov chain. This super-selection rule comes from the fact that each MemCPU coprocessor run starts with random initial conditions for voltages and currents, and therefore there could be a fluctuation in the output objective due to the finite (and short) simulation time~\cite{tuning}. Once we have the best parameter choice per Markov chain, we can run the same problem multiple times with random initial conditions for the same set of parameters and select the set of parameters that statistically arrives fastest to the best objective in a given time out. In this way, I extracted the set of parameters reported in Table~\ref{parameters}.         

\begin{figure}[t!]
	\includegraphics[width=\columnwidth]{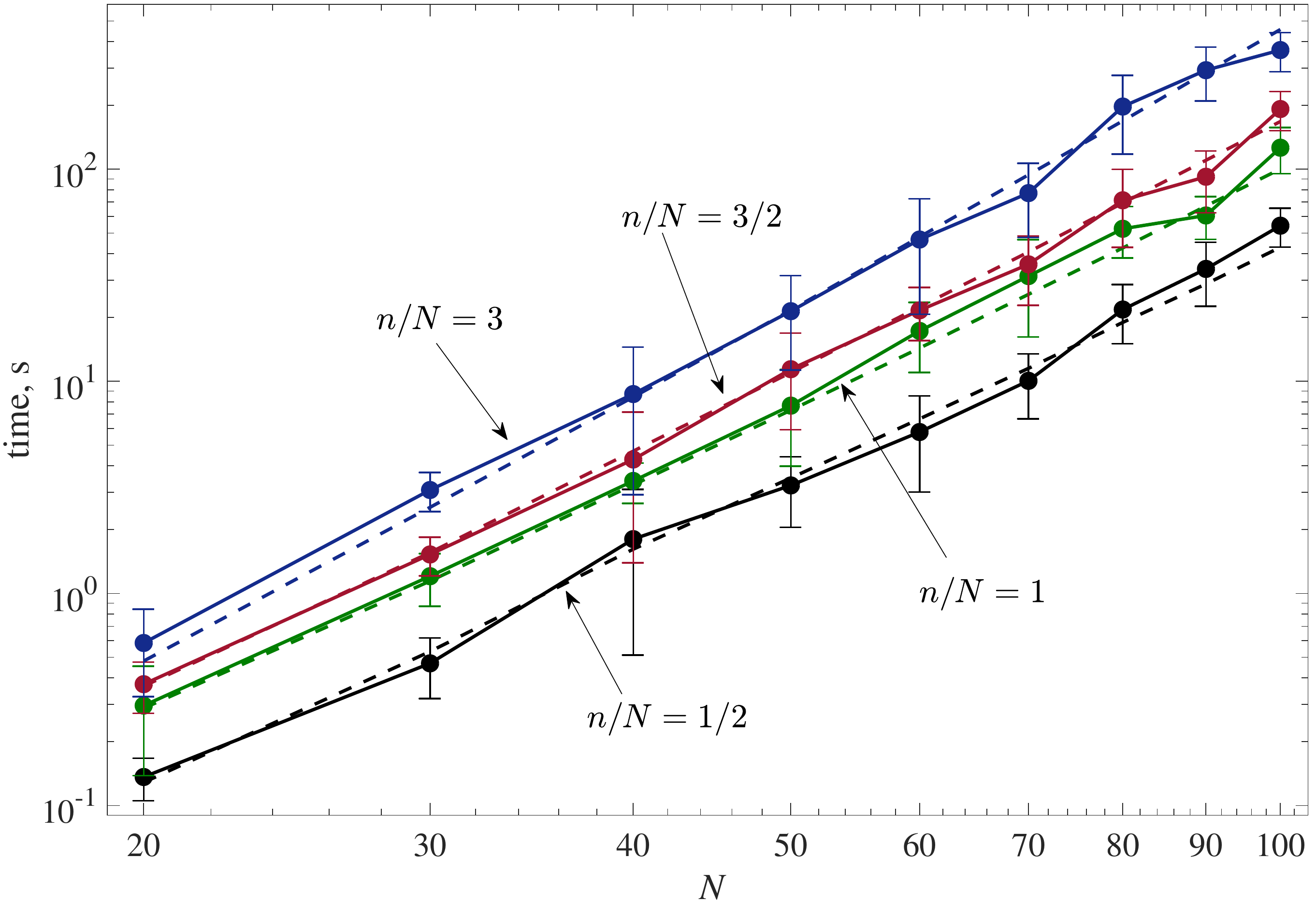}
	\caption{MemCPU coprocessor run time versus $N$ to find a configuration of containers with total mass larger that than 99.9\% of the maximum mass allowed on the aircraft or of the total payload if smaller than the maximum mass allowed. Dots connected by solid lines are the average time for 100 different ALO instances generated using the code in Fig.~\ref{matlabcode}. Dashed curves are the scaling relation~\eqref{scaling}. Runs have been carried out on an Intel(R) Xeon(R) Gold 6138 CPU @ 2.00GHz. \label{N_vs_time}}
\end{figure}

\begin{figure}[t!]
	\includegraphics[width=\columnwidth]{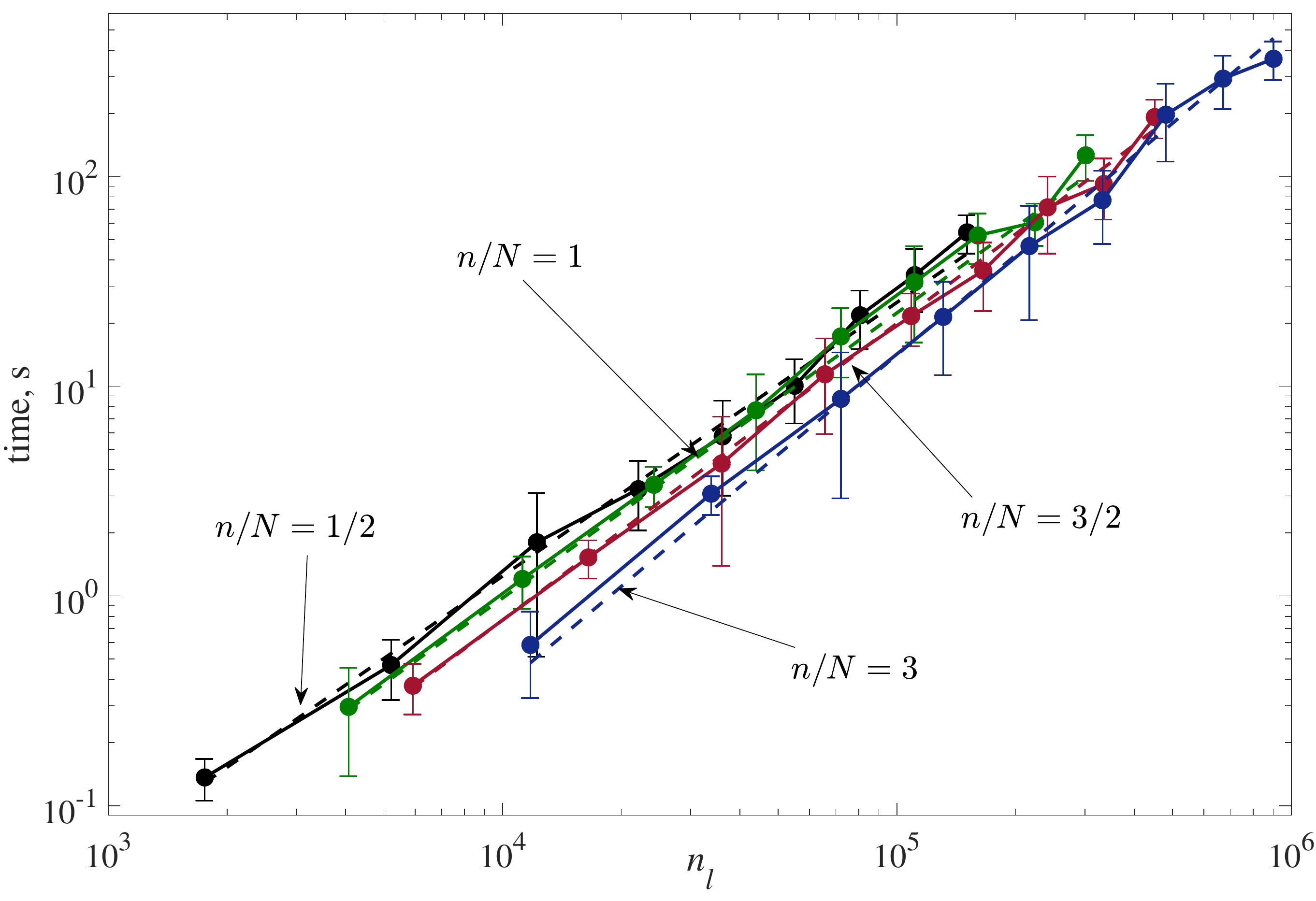}
	\caption{MemCPU coprocessor run time versus $n_l$ to find a configuration of containers with total mass larger that than 99.9\% of the maximum mass allowed on the aircraft or of the total payload if smaller than the maximum mass allowed. Dots connected by solid lines are the average time for 100 different ALO instances generated using the code in Fig.~\ref{matlabcode}. Dashed curves are the scaling relation~\eqref{scaling}. Runs have been carried out on an Intel(R) Xeon(R) Gold 6138 CPU @ 2.00GHz. \label{nl_vs_time}}
\end{figure}

Using the parameters in Table~\ref{parameters}, the MemCPU software has been tested on a set of ALO instances generated for different $n$ and $N$ using the code in Fig.~\ref{matlabcode}. The created benchmark consists of 100 ALO instances for each pair $(n,N)$. Each instance has a different container mass distribution randomly generated by the subfunction {\fontfamily{qcr}\selectfont container\_distribution\_generator} of Fig.~\ref{matlabcode}. For each instance, the container sizes were $n_1 = \dim(K_1) = n/2$,  $n_2 = \dim(K_2) = n/3$ and  $n_3 = \dim(K_3) = n/6$ with appropriate rounding, in order to have integer $n_j$ and $n_1+n_2+n_3=n$.

Since the value of the objective related to the global minimum of the ALO problem is not known and the MemCPU technology does not provide proof of optimality, I have set a very tight threshold to assess the quality of the solution: accept the solution if the objective function is smaller than $-\tau W_p$ if $W_p\leq\sum m_k$ or smaller than $-\tau \sum m_k$ otherwise, with $\tau=0.999$. This is equivalent to requiring a container selection that is larger than 99.9\% of either the maximum weight allowed on the aircraft or of the total payload if the latter is smaller than the former. In Fig.~\ref{N_vs_time} and~\ref{nl_vs_time} the time to find the solution for $\tau = 0.999$ is reported as a function of $N$ (Fig.~\ref{N_vs_time}) or as a function of the number $n_l$ of nonzero entries in the matrix defined by constraints~\eqref{constraints} (Fig.~\ref{nl_vs_time}) for different ratios $r=n/N$. 

The results as a function of $n_l$ are useful for properly assessing the scaling of the MemCPU coprocessor. Indeed, the MemCPU software associates an SOAG with a number of terminals equal to the terms in the constraint to each constraint. Each terminal includes a simulated dynamic correcting module (DCM)~\cite{Traversa2018a,DMM2}. Therefore, the number of DCMs that the MemCPU software simulates is equal to $n_l$, and it is natural to measure the complexity in terms of $n_l$ that is at the same time also a measure of the size of an IP problem. The interpolation curve on $\log$-$\log$ scale in the plane $n_l$-time provides the following relation:
\begin{equation} 
t=10^{-0.65r-4.8}n_l^{0.11r+1.25}\label{scaling}
\end{equation}
where the dependence on $r$ is valid for $0.5\leq r\leq3$. The equation~\eqref{scaling} shows a sub-quadratic scaling of the time to solution versus $n_l$. From \eqref{constraints} it is easy to realize that $n_l\propto nN^2$ and, substituting this relation in equation~\eqref{scaling}, the scaling as a function of $N$ and $n$ can be recovered.

It is worth discussing the dependence of Eq.~\eqref{scaling} on $r$. The first thing that can be noticed is that the dependence of $n_l$ on $r$ is very weak. The coefficient 0.11 in \eqref{scaling} makes the exponent range from 1.26 to 1.58 for $0.5\leq r\leq3$. On top of the weak dependence, there is compensation from the stronger and negative dependence of the prefactor $10^{-0.65r-4.8}$ on $r$. However, increasing $r$, an evident dependence of the exponent on $r$ disappears, and it saturates for $r>3$. This is not surprising behavior because for larger $r$ the problem becomes ``easier'' since there are many more choices of container selections that respect the constraints, and the burden in the calculation depends only on the size of the ALO. We do not show here numerical experiments to support the latter claim because considering $n$ much larger the $N$ seems meaningless in practical cases.     

\begin{figure}[t!]
	\includegraphics[width=\columnwidth]{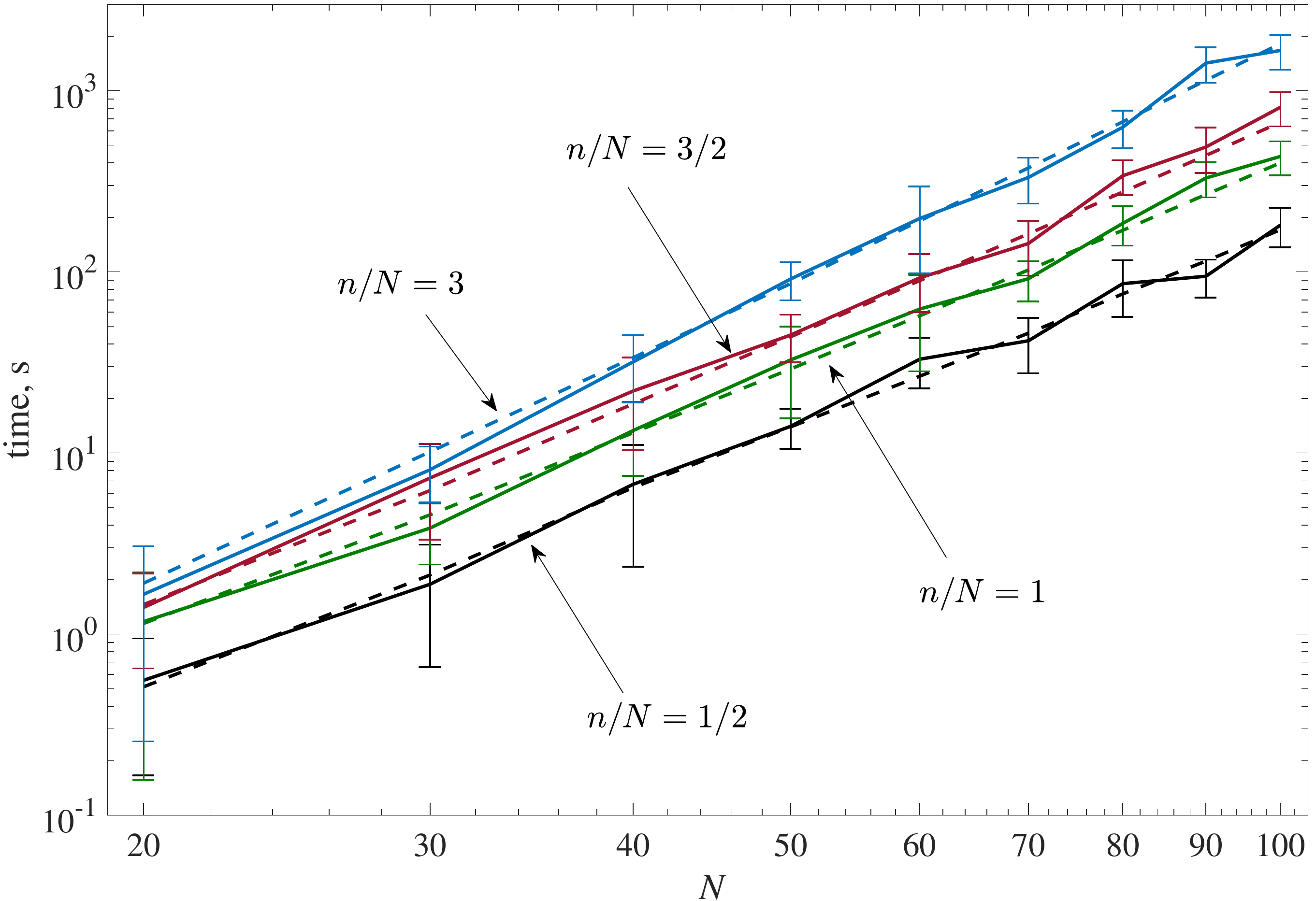}
	\caption{MemCPU coprocessor run time versus $N$ to find a configuration of containers that a) has a total mass larger that than 99.8\% of the maximum mass allowed on the aircraft or of the total payload if smaller than the maximum mass allowed and b) optimizes the center of gravity of the aircraft. Dots connected by solid lines are the average time for 100 different ALO instances generated using the code in Fig.~\ref{matlabcode}. Dashed curves are the scaling relation~\eqref{scaling}. Runs have been carried out on an Intel(R) Xeon(R) Gold 6138 CPU @ 2.00GHz. \label{N_vs_time_xmin}}
\end{figure}
Let us consider now the ALO problem described in (section~\ref{CG_optimization_section}). In order to handle this problem by using a non-modified version of MemCPU software, I have created a sequence of linear IP problems that converges to the solution of the non linear problem of (section~\ref{CG_optimization_section}). This sequence of problems can be easily created making use of the code of Fig.~\ref{matlabcode} as follows: 
\begin{enumerate}[a.] 
	\item Set $x^{\min}_{cg}$ and $x^{\max}_{cg}$ for some initial values, for example, the ones given in Table~\ref{Airbusdataset}. 
	\item Generate the problem for a given $x^{\min}_{cg}$ and $x^{\max}_{cg}$.
	\item Add to the matrix of the constraints the objective function of the problem as an extra constraint. For the right hand side of this constraint set $-\tau W^{\max}$ where $W^{\max}=W_p$ if $W_p<\sum m_k$ or  $W^{\max}=\sum m_k$ otherwise. 
	\item Substitute the objective function with a null objective function and solve the problem. 
	\item If a solution to the problem is found, evaluate $x_{cg}$ by means of {\fontfamily{qcr}\selectfont J}, {\fontfamily{qcr}\selectfont VL}, {\fontfamily{qcr}\selectfont V}, {\fontfamily{qcr}\selectfont xe\_cg} and {\fontfamily{qcr}\selectfont We} as reported in the code of figure~\ref{matlabcode}
	\item if $x_{cg}<x^t_{cg}$ set $x^{\min}_{cg}=x_{cg}+\epsilon$ otherwise set $x^{\max}_{cg}=x_{cg}-\epsilon$ for some $\epsilon>0$. Go back to b.   
\end{enumerate}

\begin{figure}[t!]
	\includegraphics[width=\columnwidth]{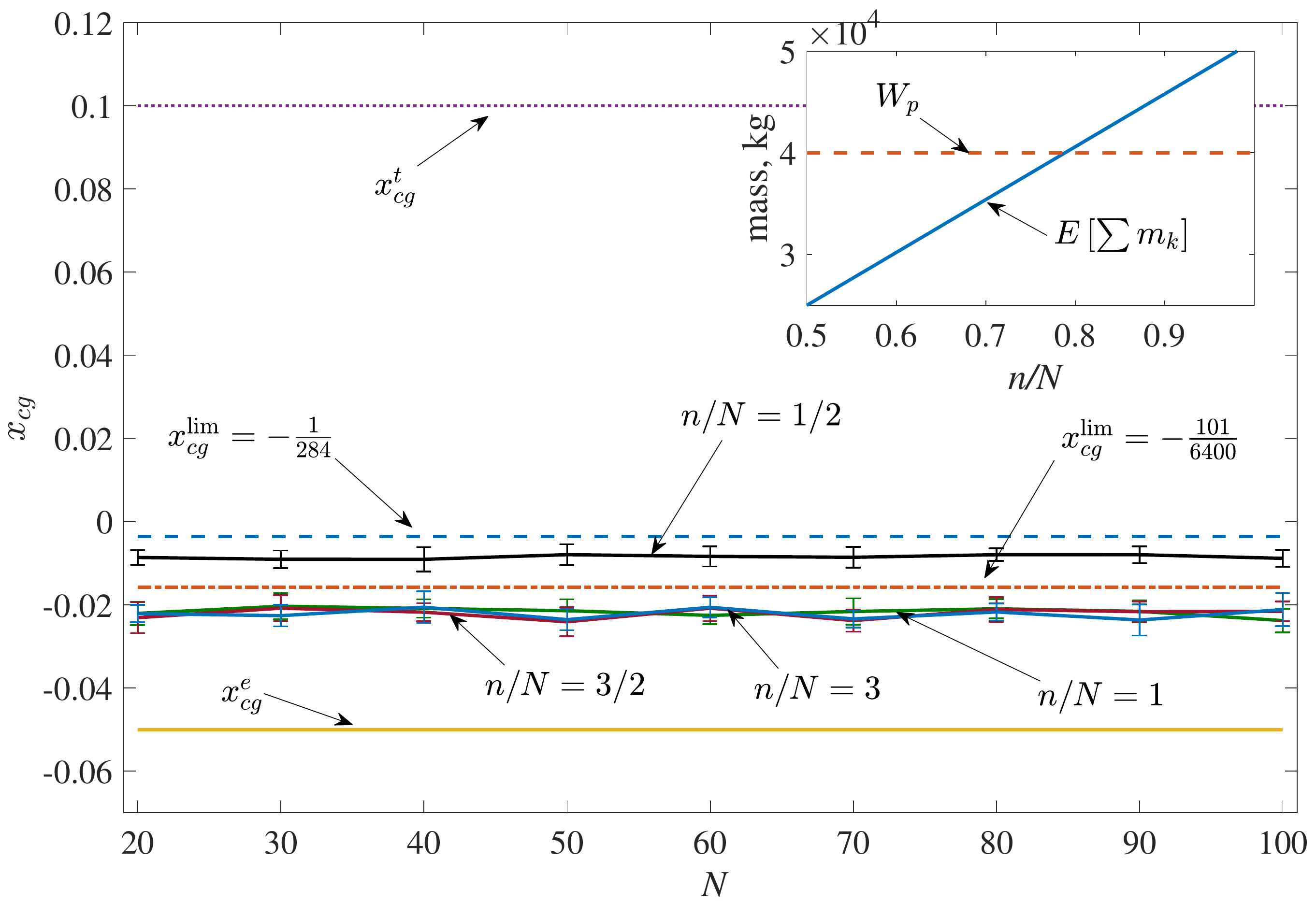}
	\caption{Center of gravity of the system aircraft + selected payload with total carried mass larger than $99.8\%$ of $W^{\max} = \min(W_p,\sum m_k)$. In the inset, the mean of the total mass generated by the subfunction {\fontfamily{qcr}\selectfont container\_distribution\_generator} of Fig.~\ref{matlabcode} and $W_p$ versus $r=n/N$ is plotted. \label{xcg_vs_time}}
\end{figure}

In Fig.~\ref{N_vs_time_xmin} the time to solution is reported for selecting the payload that maximizes the mass ($tau=0.998$ has been used in this case) and at the same time provides the center of gravity as close as possible to the target, for different values of $N$ and $n$. The dashed curves that interpolate this version of the ALO result as a shift of the equation~\eqref{scaling} where the coefficient $10^{-0.65r-4.8}$ is replaced by $10^{-0.65r-4.2}$. Finding the same scaling properties as in the other problem is not surprising, since solving this ALO problem follows the steps a.-f., and so is a cascade of problems similar to the ones solved in Fig.~\ref{N_vs_time}.
  
Finally, it is worth noticing where the optimized centers of gravity are located. In Fig.~\ref{xcg_vs_time}, we can see that for $n=N/2$ the centers of gravity approach the first limit of Fig.~\ref{xlimitcg}. This is not surprising in this case because of the distribution of containers, $W^{\max}\approx W_p/2$ (see inset of Fig.~\ref{xcg_vs_time}). On the other hand, if $n\ge N$, $W^{\max}> W_p$ (see inset of Fig.~\ref{xcg_vs_time}) and the optimized centers of gravity approach the second limit of Fig.~\ref{xlimitcg}.

\subsection{Further improvements}\label{improvements}

We briefly discuss here a few improvements that can be made in order to have an even more efficient time-to-solution for the Airbus ALO problem computed through a MemCPU coprocessor. However, since they do not qualitatively affect the scaling assessed in this work, the implementation of these is beyond the current scope. 

The first improvement concerns the initial values of the objective function~\eqref{objective}. The MemCPU coprocessor searches solutions that have an objective equal to or lower than a target. Once found it decreases the objective and searches for the next solution (see section~\ref{Mem_section} and~\cite{Traversa2018a}). In this work I have used infinite as the initial objective. With this choice, the MemCPU software provides many solutions at lower and lower objectives, progressing from the larger to smaller values. Even though this convergence is usually fast, this can be improved even more by giving as a starting value for the objective function $\tau \min(W_p,\sum m_k)$.   

Another improvement that scales the time to solution down by a few orders of magnitude is running the MemCPU coprocessor on GPUs rather than CPUs~\cite{SharpGPU}. In fact, the computation performed by MemCPU coprocessor is nothing other than a circuit simulation that can efficiently be distributed on GPUs~\cite{SharpGPU}.  

Finally, the search for maximal carried freight that optimizes the center of gravity at the same time can be accelerated by directly implementing the nonlinear objective function~\eqref{newobjective}, avoiding the sequence of ALO problems discussed in section~\ref{MemCPU_section} and solving only one ALO problem.   

\section{Conclusions}\label{Conclusions}

In summary, I have shown how to employ MemComputing through the the MemCPU coprocessor to tackle the 5th Airbus problem efficiently without a quantum computer.

In order to use the the MemCPU coprocessor for the 5th Airbus problem, which is an aircraft loading optimization problem, an IP problem has been formulated that includes all constraints required by Airbus with no exception and no approximation.  

The scaling properties assessed in this work show sub-quadratic scaling by the MemCPU coprocessor as a function of the size of the problem measured by the number of nonzero elements of the constraint matrix of the associated IP problem. 

The scaling properties of the MemCPU coprocessor allow efficient solutions of large ALO problems and it represents a solution ready to be deployed in the field.  

Finally, I have also discussed extra improvements that can be made in order to further (and substantially) accelerate the time to solution for the ALO problem by the MemCPU coprocessor.

\section{Acknowledgments}\label{Acknowledgments}

I thank the MemComputing team for useful discussions and in particular Dr. Tristan Sharp and John Beane for the revision of the manuscript. This work has been supported by MemComputing, Inc. 

\section{References}\label{References}

%\bibliography{IEEEabrv,IEEEexample}
\bibliographystyle{naturemag}
\bibliography{SUSYref}

\begin{thebibliography}{10}
\expandafter\ifx\csname url\endcsname\relax
  \def\url#1{\texttt{#1}}\fi
\expandafter\ifx\csname urlprefix\endcsname\relax\def\urlprefix{URL }\fi
\providecommand{\bibinfo}[2]{#2}
\providecommand{\eprint}[2][]{\url{#2}}

\bibitem{Eiben2015}
\bibinfo{author}{Eiben, A.} \& \bibinfo{author}{Smith, J.}
\newblock \emph{\bibinfo{title}{Introduction to Evolutionary Computing}}
  (\bibinfo{publisher}{Springer Berlin Heidelberg}, \bibinfo{year}{2015}).

\bibitem{Kephart2003}
\bibinfo{author}{Kephart, J.} \& \bibinfo{author}{Chess, D.}
\newblock \bibinfo{title}{The vision of autonomic computing}.
\newblock \emph{\bibinfo{journal}{Computer}} \textbf{\bibinfo{volume}{36}},
  \bibinfo{pages}{41--50} (\bibinfo{year}{2003}).

\bibitem{Vestias2014}
\bibinfo{author}{Vestias, M.} \& \bibinfo{author}{Neto, H.}
\newblock \bibinfo{title}{Trends of {CPU}, {GPU} and {FPGA} for
  high-performance computing}.
\newblock In \emph{\bibinfo{booktitle}{2014 24th International Conference on
  Field Programmable Logic and Applications ({FPL})}}
  (\bibinfo{publisher}{{IEEE}}, \bibinfo{year}{2014}).

\bibitem{Hashem2015}
\bibinfo{author}{Hashem, I. A.~T.} \emph{et~al.}
\newblock \bibinfo{title}{The rise of {\textquotedblleft}big
  data{\textquotedblright} on cloud computing: Review and open research
  issues}.
\newblock \emph{\bibinfo{journal}{Information Systems}}
  \textbf{\bibinfo{volume}{47}}, \bibinfo{pages}{98--115}
  (\bibinfo{year}{2015}).

\bibitem{Burr2016}
\bibinfo{author}{Burr, G.~W.} \emph{et~al.}
\newblock \bibinfo{title}{Neuromorphic computing using non-volatile memory}.
\newblock \emph{\bibinfo{journal}{Advances in Physics: X}}
  \textbf{\bibinfo{volume}{2}}, \bibinfo{pages}{89--124}
  (\bibinfo{year}{2016}).

\bibitem{Torrejon2017}
\bibinfo{author}{Torrejon, J.} \emph{et~al.}
\newblock \bibinfo{title}{Neuromorphic computing with nanoscale spintronic
  oscillators}.
\newblock \emph{\bibinfo{journal}{Nature}} \textbf{\bibinfo{volume}{547}},
  \bibinfo{pages}{428--431} (\bibinfo{year}{2017}).

\bibitem{Furber2014}
\bibinfo{author}{Furber, S.~B.}, \bibinfo{author}{Galluppi, F.},
  \bibinfo{author}{Temple, S.} \& \bibinfo{author}{Plana, L.~A.}
\newblock \bibinfo{title}{The {SpiNNaker} project}.
\newblock \emph{\bibinfo{journal}{Proceedings of the {IEEE}}}
  \textbf{\bibinfo{volume}{102}}, \bibinfo{pages}{652--665}
  (\bibinfo{year}{2014}).

\bibitem{Linke2017}
\bibinfo{author}{Linke, N.~M.} \emph{et~al.}
\newblock \bibinfo{title}{Experimental comparison of two quantum computing
  architectures}.
\newblock \emph{\bibinfo{journal}{Proceedings of the National Academy of
  Sciences}} \textbf{\bibinfo{volume}{114}}, \bibinfo{pages}{3305--3310}
  (\bibinfo{year}{2017}).

\bibitem{Bennett2000}
\bibinfo{author}{Bennett, C.~H.} \& \bibinfo{author}{DiVincenzo, D.~P.}
\newblock \bibinfo{title}{Quantum information and computation}.
\newblock \emph{\bibinfo{journal}{Nature}} \textbf{\bibinfo{volume}{404}},
  \bibinfo{pages}{247--255} (\bibinfo{year}{2000}).

\bibitem{Ladd2010}
\bibinfo{author}{Ladd, T.~D.} \emph{et~al.}
\newblock \bibinfo{title}{Quantum computers}.
\newblock \emph{\bibinfo{journal}{Nature}} \textbf{\bibinfo{volume}{464}},
  \bibinfo{pages}{45--53} (\bibinfo{year}{2010}).

\bibitem{Raha2007}
\bibinfo{author}{Raha, K.} \emph{et~al.}
\newblock \bibinfo{title}{The role of quantum mechanics in structure-based drug
  design}.
\newblock \emph{\bibinfo{journal}{Drug Discovery Today}}
  \textbf{\bibinfo{volume}{12}}, \bibinfo{pages}{725--731}
  (\bibinfo{year}{2007}).

\bibitem{Harrow2017}
\bibinfo{author}{Harrow, A.~W.} \& \bibinfo{author}{Montanaro, A.}
\newblock \bibinfo{title}{Quantum computational supremacy}.
\newblock \emph{\bibinfo{journal}{Nature}} \textbf{\bibinfo{volume}{549}},
  \bibinfo{pages}{203--209} (\bibinfo{year}{2017}).

\bibitem{MITtechrev}
\bibinfo{title}{\href{https://www.technologyreview.com/s/612760/quantum-computers-component-shortage/}{MIT
  Technology Review}}.

\bibitem{Dyakonov2019}
\bibinfo{author}{Dyakonov, M.}
\newblock \bibinfo{title}{When will useful quantum computers be constructed?
  not in the foreseeable future, this physicist argues. here's why: The case
  against: Quantum computing}.
\newblock \emph{\bibinfo{journal}{{IEEE} Spectrum}}
  \textbf{\bibinfo{volume}{56}}, \bibinfo{pages}{24--29}
  (\bibinfo{year}{2019}).

\bibitem{Tang2018}
\bibinfo{author}{Tang, E.}
\newblock \bibinfo{title}{A quantum-inspired classical algorithm for
  recommendation systems}.
\newblock \emph{\bibinfo{journal}{arXiv:1807.04271}}  (\bibinfo{year}{2018}).
\newblock \eprint{http://arxiv.org/abs/1807.04271v1}.

\bibitem{Accenture_report}
\bibinfo{title}{\href{https://www.accenture.com/t20190227T065754Z__w__/us-en/_acnmedia/PDF-94/Accenture-TechVision-2019-Tech-Trends-Report.pdf}{Accenture
  report}}.

\bibitem{AribusURL}
\bibinfo{title}{\href{https://www.airbus.com/newsroom/press-releases/en/2019/01/airbus-launches-quantum-computing-challenge-to-transform-the-aircraft-lifecycle.html}{Airbus
  challenge press-release};
  \href{https://www.airbus.com/innovation/airbus-quantum-computing-challenge.html}{Airbus
  challenge web-page}}.

\bibitem{Schrijver1998}
\bibinfo{author}{Schrijver}.
\newblock \emph{\bibinfo{title}{Theory of Linear Integer Programming}}
  (\bibinfo{publisher}{John Wiley \& Sons}, \bibinfo{year}{1998}).

\bibitem{complexity_bible}
\bibinfo{author}{Garey, M.~R.} \& \bibinfo{author}{Johnson, D.~S.}
\newblock \emph{\bibinfo{title}{Computers and Intractability; A Guide to the
  Theory of NP-Completeness}} (\bibinfo{publisher}{W. H. Freeman \& Co.},
  \bibinfo{address}{New York, NY, USA}, \bibinfo{year}{1990}).

\bibitem{GLPK}
\bibinfo{title}{Gnu linear programming kit, version 4.32}.
\newblock \urlprefix\url{http://www.gnu.org/software/glpk/glpk.html}.

\bibitem{Forrest2018}
\bibinfo{author}{Forrest, J.} \emph{et~al.}
\newblock \bibinfo{title}{Coin-or/cbc: Version 2.9.9} (\bibinfo{year}{2018}).

\bibitem{Gleixner2018}
\bibinfo{author}{Gleixner, A.} \emph{et~al.}
\newblock \bibinfo{title}{{The SCIP Optimization Suite 6.0}}.
\newblock \bibinfo{type}{ZIB-Report} \bibinfo{number}{18-26},
  \bibinfo{institution}{Zuse Institute Berlin} (\bibinfo{year}{2018}).
\newblock \urlprefix\url{http://nbn-resolving.de/urn:nbn:de:0297-zib-69361}.

\bibitem{Ralphs2017}
\bibinfo{author}{Ralphs, T.} \emph{et~al.}
\newblock \bibinfo{title}{Coin-or/symphony: Version 5.6.16}
  (\bibinfo{year}{2017}).

\bibitem{Berkelaar2004}
\bibinfo{author}{Berkelaar, M.}, \bibinfo{author}{Eikland, K.} \&
  \bibinfo{author}{Notebaert, P.}
\newblock \bibinfo{title}{{lp\_solve} 5.5, open source (mixed-integer) linear
  programming system}.
\newblock \bibinfo{howpublished}{Software} (\bibinfo{year}{2004}).
\newblock \urlprefix\url{http://lpsolve.sourceforge.net/5.5/}.

\bibitem{Gurobi}
\urlprefix\url{http://www.gurobi.com}.

\bibitem{CPLEX}
\urlprefix\url{https://www.ibm.com/analytics/cplex-optimizer}.

\bibitem{FICO}
\urlprefix\url{http://www.fico.com/en/products/fico-xpress-optimization}.

\bibitem{MATLAB}
\urlprefix\url{http://mathworks.com}.

\bibitem{applegate2006concorde}
\bibinfo{author}{Applegate, D.}, \bibinfo{author}{Bixby, R.},
  \bibinfo{author}{Chvatal, V.} \& \bibinfo{author}{Cook, W.}
\newblock \bibinfo{title}{Concorde tsp solver} (\bibinfo{year}{2006}).

\bibitem{Floudas2005}
\bibinfo{author}{Floudas, C.~A.} \& \bibinfo{author}{Lin, X.}
\newblock \bibinfo{title}{Mixed integer linear programming in process
  scheduling: Modeling, algorithms, and applications}.
\newblock \emph{\bibinfo{journal}{Annals of Operations Research}}
  \textbf{\bibinfo{volume}{139}}, \bibinfo{pages}{131--162}
  (\bibinfo{year}{2005}).

\bibitem{Boston99}
\bibinfo{author}{Boston, K.} \& \bibinfo{author}{Bettinger, P.}
\newblock \bibinfo{title}{An analysis of monte carlo integer programming,
  simulated annealing, and tabu search heuristics for solving spatial harvest
  scheduling problems}.
\newblock \emph{\bibinfo{journal}{Forest Science}}
  \textbf{\bibinfo{volume}{45}}, \bibinfo{pages}{292--301}
  (\bibinfo{year}{1999}).

\bibitem{sorensen2014hybridizing}
\bibinfo{author}{S{\o}rensen, M.} \& \bibinfo{author}{Stidsen, T.~R.}
\newblock \bibinfo{title}{Hybridizing integer programming and metaheuristics
  for solving high school timetabling}.
\newblock In \emph{\bibinfo{booktitle}{Proceedings of the 10th international
  conference of the practice and theory of automated timetabling}},
  \bibinfo{pages}{557--560} (\bibinfo{year}{2014}).

\bibitem{Danna2004}
\bibinfo{author}{Danna, E.}, \bibinfo{author}{Rothberg, E.} \&
  \bibinfo{author}{Pape, C.~L.}
\newblock \bibinfo{title}{Exploring relaxation induced neighborhoods to improve
  {MIP} solutions}.
\newblock \emph{\bibinfo{journal}{Mathematical Programming}}
  \textbf{\bibinfo{volume}{102}}, \bibinfo{pages}{71--90}
  (\bibinfo{year}{2004}).

\bibitem{Glover1977}
\bibinfo{author}{Glover, F.}
\newblock \bibinfo{title}{Heuristics for integer programming using surrogate
  constraints}.
\newblock \emph{\bibinfo{journal}{Decision Sciences}}
  \textbf{\bibinfo{volume}{8}}, \bibinfo{pages}{156--166}
  (\bibinfo{year}{1977}).

\bibitem{Traversa2018a}
\bibinfo{author}{Traversa, F.~L.} \& \bibinfo{author}{{Di Ventra}, M.}
\newblock \bibinfo{title}{Memcomputing integer linear programming}.
\newblock \emph{\bibinfo{journal}{arXiv:1808.09999}}  (\bibinfo{year}{2018}).
\newblock \eprint{http://arxiv.org/abs/1808.09999v1}.

\bibitem{barnhart1998branch}
\bibinfo{author}{Barnhart, C.}, \bibinfo{author}{Johnson, E.~L.},
  \bibinfo{author}{Nemhauser, G.~L.}, \bibinfo{author}{Savelsbergh, M.~W.} \&
  \bibinfo{author}{Vance, P.~H.}
\newblock \bibinfo{title}{Branch-and-price: Column generation for solving huge
  integer programs}.
\newblock \emph{\bibinfo{journal}{Operations research}}
  \textbf{\bibinfo{volume}{46}}, \bibinfo{pages}{316--329}
  (\bibinfo{year}{1998}).

\bibitem{benders1962partitioning}
\bibinfo{author}{Benders, J.~F.}
\newblock \bibinfo{title}{Partitioning procedures for solving mixed-variables
  programming problems}.
\newblock \emph{\bibinfo{journal}{Numerische mathematik}}
  \textbf{\bibinfo{volume}{4}}, \bibinfo{pages}{238--252}
  (\bibinfo{year}{1962}).

\bibitem{hooker2003logic}
\bibinfo{author}{Hooker, J.~N.} \& \bibinfo{author}{Ottosson, G.}
\newblock \bibinfo{title}{Logic-based benders decomposition}.
\newblock \emph{\bibinfo{journal}{Mathematical Programming}}
  \textbf{\bibinfo{volume}{96}}, \bibinfo{pages}{33--60}
  (\bibinfo{year}{2003}).

\bibitem{abara1989applying}
\bibinfo{author}{Abara, J.}
\newblock \bibinfo{title}{Applying integer linear programming to the fleet
  assignment problem}.
\newblock \emph{\bibinfo{journal}{Interfaces}} \textbf{\bibinfo{volume}{19}},
  \bibinfo{pages}{20--28} (\bibinfo{year}{1989}).

\bibitem{kroon2009new}
\bibinfo{author}{Kroon, L.} \emph{et~al.}
\newblock \bibinfo{title}{The new dutch timetable: The or revolution}.
\newblock \emph{\bibinfo{journal}{Interfaces}} \textbf{\bibinfo{volume}{39}},
  \bibinfo{pages}{6--17} (\bibinfo{year}{2009}).

\bibitem{stahlbock2008operations}
\bibinfo{author}{Stahlbock, R.} \& \bibinfo{author}{Vo{\ss}, S.}
\newblock \bibinfo{title}{Operations research at container terminals: a
  literature update}.
\newblock \emph{\bibinfo{journal}{OR spectrum}} \textbf{\bibinfo{volume}{30}},
  \bibinfo{pages}{1--52} (\bibinfo{year}{2008}).

\bibitem{melo2009facility}
\bibinfo{author}{Melo, M.~T.}, \bibinfo{author}{Nickel, S.} \&
  \bibinfo{author}{Saldanha-Da-Gama, F.}
\newblock \bibinfo{title}{Facility location and supply chain management--a
  review}.
\newblock \emph{\bibinfo{journal}{European journal of operational research}}
  \textbf{\bibinfo{volume}{196}}, \bibinfo{pages}{401--412}
  (\bibinfo{year}{2009}).

\bibitem{bard2003staff}
\bibinfo{author}{Bard, J.~F.}, \bibinfo{author}{Binici, C.} \emph{et~al.}
\newblock \bibinfo{title}{Staff scheduling at the united states postal
  service}.
\newblock \emph{\bibinfo{journal}{Computers \& Operations Research}}
  \textbf{\bibinfo{volume}{30}}, \bibinfo{pages}{745--771}
  (\bibinfo{year}{2003}).

\bibitem{UMM}
\bibinfo{author}{Traversa, F.~L.} \& \bibinfo{author}{{Di Ventra}, M.}
\newblock \bibinfo{title}{Universal memcomputing machines}.
\newblock \emph{\bibinfo{journal}{IEEE Trans. Neural Netw. Learn. Syst.}}
  \textbf{\bibinfo{volume}{26}}, \bibinfo{pages}{2702} (\bibinfo{year}{2015}).

\bibitem{Di_Ventra2018}
\bibinfo{author}{Di~Ventra, M.} \& \bibinfo{author}{Traversa, F.~L.}
\newblock \bibinfo{title}{Perspective: Memcomputing: Leveraging memory and
  physics to compute efficiently}.
\newblock \emph{\bibinfo{journal}{Journal of Applied Physics}}
  \textbf{\bibinfo{volume}{123}}, \bibinfo{pages}{180901}
  (\bibinfo{year}{2018}).

\bibitem{DMM2}
\bibinfo{author}{Traversa, F.~L.} \& \bibinfo{author}{{Di Ventra}, M.}
\newblock \bibinfo{title}{Polynomial-time solution of prime factorization and
  np-complete problems with digital memcomputing machines}.
\newblock \emph{\bibinfo{journal}{Chaos: An Interdisciplinary Journal of
  Nonlinear Science}} \textbf{\bibinfo{volume}{27}}, \bibinfo{pages}{023107}
  (\bibinfo{year}{2017}).

\bibitem{no-chaos}
\bibinfo{author}{Di~Ventra, M.} \& \bibinfo{author}{Traversa, F.~L.}
\newblock \bibinfo{title}{Absence of chaos in digital memcomputing machines
  with solutions}.
\newblock \emph{\bibinfo{journal}{Phys. Lett. A}}
  \textbf{\bibinfo{volume}{381}}, \bibinfo{pages}{3255} (\bibinfo{year}{2017}).

\bibitem{noperiod}
\bibinfo{author}{{Di Ventra}, M.} \& \bibinfo{author}{Traversa, F.~L.}
\newblock \bibinfo{title}{Absence of periodic orbits in digital memcomputing
  machines with solutions}.
\newblock \emph{\bibinfo{journal}{Chaos: An Interdisciplinary Journal of
  Nonlinear Science}} \textbf{\bibinfo{volume}{27}}, \bibinfo{pages}{101101}
  (\bibinfo{year}{2017}).

\bibitem{MemWeb}
\urlprefix\url{www.memcpu.com}.

\bibitem{patentSOLC}
\bibinfo{author}{{Di Ventra}, M.} \& \bibinfo{author}{Traversa, F.~L.}
\newblock \bibinfo{title}{Self-organizing logic gates and circuits and complex
  problem solving with self-organizing logic circuits, {US} patent application
  {No}. 15/557,641, {US} patent {No}. 9,911,080} (\bibinfo{year}{2018}).

\bibitem{traversaNP}
\bibinfo{author}{Traversa, F.~L.}, \bibinfo{author}{Ramella, C.},
  \bibinfo{author}{Bonani, F.} \& \bibinfo{author}{{Di Ventra}, M.}
\newblock \bibinfo{title}{Memcomputing {NP}-complete problems in polynomial
  time using polynomial resources and collective states}.
\newblock \emph{\bibinfo{journal}{Science Advances}}
  \textbf{\bibinfo{volume}{1}}, \bibinfo{pages}{e1500031}
  (\bibinfo{year}{2015}).

\bibitem{collective}
\bibinfo{author}{Traversa, F.~L.}
\newblock \bibinfo{title}{Collective {Computing}}.
\newblock \emph{\bibinfo{journal}{In preparation}}  (\bibinfo{year}{2018}).

\bibitem{topo}
\bibinfo{author}{Di~Ventra, M.}, \bibinfo{author}{Traversa, F.~L.} \&
  \bibinfo{author}{Ovchinnikov, I.~V.}
\newblock \bibinfo{title}{Topological field theory and computing with
  instantons}.
\newblock \emph{\bibinfo{journal}{Ann. Phys. (Berlin)}}
  \bibinfo{pages}{1700123} (\bibinfo{year}{2017}).

\bibitem{Bearden2018}
\bibinfo{author}{Bearden, S.~R.}, \bibinfo{author}{Manukian, H.},
  \bibinfo{author}{Traversa, F.~L.} \& \bibinfo{author}{{Di Ventra}, M.}
\newblock \bibinfo{title}{Instantons in self-organizing logic gates}.
\newblock \emph{\bibinfo{journal}{Physical Review Applied}}
  \textbf{\bibinfo{volume}{9}} (\bibinfo{year}{2018}).

\bibitem{Traversa2018}
\bibinfo{author}{Traversa, F.~L.}, \bibinfo{author}{Cicotti, P.},
  \bibinfo{author}{Sheldon, F.} \& \bibinfo{author}{{Di Ventra}, M.}
\newblock \bibinfo{title}{Evidence of exponential speed-up in the solution of
  hard optimization problems}.
\newblock \emph{\bibinfo{journal}{Complexity}} \textbf{\bibinfo{volume}{2018}},
  \bibinfo{pages}{1--13} (\bibinfo{year}{2018}).

\bibitem{Sheldon2018}
\bibinfo{author}{Sheldon, F.}, \bibinfo{author}{Cicotti, P.},
  \bibinfo{author}{Traversa, F.~L.} \& \bibinfo{author}{{Di Ventra}, M.}
\newblock \bibinfo{title}{Stress-testing memcomputing on hard combinatorial
  optimization problems}.
\newblock \emph{\bibinfo{journal}{Preprint arXiv:1807.00107}}
  (\bibinfo{year}{2018}).
\newblock \eprint{http://arxiv.org/abs/1807.00107v1}.

\bibitem{spinglass}
\bibinfo{author}{Sheldon, F.}, \bibinfo{author}{Traversa, F.~L.} \&
  \bibinfo{author}{{Di Ventra}, M.}
\newblock \bibinfo{title}{Taming a non-convex landscape with long-range order}.
\newblock \emph{\bibinfo{journal}{In preparation}} .

\bibitem{AcceleratingDL}
\bibinfo{author}{Manukian, H.}, \bibinfo{author}{Traversa, F.~L.} \&
  \bibinfo{author}{{Di Ventra}, M.}
\newblock \bibinfo{title}{Accelerating deep learning with memcomputing}.
\newblock \emph{\bibinfo{journal}{Neural Networks}}
  \textbf{\bibinfo{volume}{110}}, \bibinfo{pages}{1--7} (\bibinfo{year}{2019}).

\bibitem{LuongMatlab}
\bibinfo{title}{\href{https://www.mathworks.com/matlabcentral/fileexchange/19618-mps-format-exporting-too}{https://www.mathworks.com/matlabcentral/fileexchange
  /19618-mps-format-exporting-tool}}.

\bibitem{tuning}
\bibinfo{author}{Pederson, E.}, \bibinfo{author}{Foertsch, J.},
  \bibinfo{author}{Qian, Z.} \& \bibinfo{author}{Traversa, F.~L.}
\newblock \bibinfo{title}{Tuning and prediction of optimal parameters for
  algorithm configuration}.
\newblock \emph{\bibinfo{journal}{In preparation}}  (\bibinfo{year}{2019}).

\bibitem{SharpGPU}
\bibinfo{author}{Sharp, T.}, , \bibinfo{author}{Qian, Z.} \&
  \bibinfo{author}{Traversa, F.~L.}
\newblock \bibinfo{title}{Distributing {Memcomputing} on {Graphics}
  {Processing} {Units}}.
\newblock \emph{\bibinfo{journal}{In preparation}}  (\bibinfo{year}{2019}).

\end{thebibliography}
%\clearpage
\begin{figure*}[t!]
\section*{Appendix}\label{Appendix}
	\begin{minipage}[t]{2\textwidth}
		{\fontfamily{qcr}\selectfont
			{\color{Blu}function} [problem, J, V, VL, xe\_cg, We, input] = ...\\
			\phantom{x}\hspace{4cm}    Airbus\_problem\_generator(xmin\_cg,xmax\_cg,n1,n2,n3,N)\\
			\\
			{\color{Green} \% airbus 5th problem generator}\\
			\\
			n = n1+n2+n3; {\color{Green}\% total number of containers}\\
			Wp = 40000; {\color{Green}\% Max payload}\\
			We = 120000; {\color{Green}\% Aircraft weight}\\
			xe\_cg = -0.05; {\color{Green}\% Center of gravity position of the aircraft without payload}\\
			Smax\_0 = 22000; {\color{Green}\% Max shear}\\
			\\		
			{\color{Green} \% generate distributions of container mass}\\
			\\
			input = container\_distribution\_generator(n1,n2,n3,N);\\
			\\
			type1 = find(input(:,2)==1);\\
			type2 = find(input(:,2)==2);\\
			type3 = find(input(:,2)==3);\\
			\\
			indexM = zeros(n,N);\\
			var = 0;\\
			{\color{Blu}for} j=1:n1\\
			\phantom{x}\hspace{12pt}indexM(type1(j),:) = var+(1:N);\\
			\phantom{x}\hspace{12pt}var = var+N;\\
			{\color{Blu}end}\\
			{\color{Blu}for} j=1:n2\\
			\phantom{x}\hspace{12pt}indexM(type2(j),:) = var+(1:N);\\
			\phantom{x}\hspace{12pt}var = var+N;\\
			{\color{Blu}end}\\
			{\color{Blu}for} j=1:n3\\
			\phantom{x}\hspace{12pt}indexM(type3(j),1:end-1) = var+(1:N-1);\\
			\phantom{x}\hspace{12pt}var = var+N-1;\\
			{\color{Blu}end}\\
			\\
			{\color{Green}\% distance defined in \eqref{length}}\\
			\\
			d = zeros(n,N);\\
			d(type1,:) = repmat(linspace(-N/2+1/2,N/2-1/2,N)/N,n1,1);\\
			d(type2,:) = repmat(linspace(-N/2+1/2,N/2-1/2,N)/N,n2,1);\\
			d(type3,1:end-1) = repmat(linspace(-N/2+1,N/2-1,N-1)/N,n3,1);\\
			\\
			{\color{Green}\% initialize matrices for IP}\\
			\\
			Aineq = [];\\
			bineq = [];\\
			\\
			{\color{Green}\% constraints on containers Eq.~(\ref{constraints}a)}\\
			\\
			Aineq = [Aineq;\\
			sparse(repmat((1:n1).',1,N),indexM(type1,:),1,n1,var);\\
			sparse(repmat((1:n2).',1,N),indexM(type2,:),1,n2,var);\\
			sparse(repmat((1:n3).',1,N-1),indexM(type3,1:end-1),1,n3,var)];\\
			\\
			bineq = [bineq; ones(n1+n2+n3,1)];\\
			\\
			\\
			\\\\}
	\end{minipage}
\end{figure*}
\begin{figure*}[htb]
	\begin{minipage}[t]{2\textwidth}
		{\fontfamily{qcr}\selectfont
			{\color{Green}\% constraints on bins Eq.~(\ref{constraints}b)}\\
			\\
			Aineq = [Aineq;\\
			sparse(1,indexM([type1; type2; type3],1),[ones(n1,1); ones(n2,1)/2; ones(n3,1)],1,...\\
			\phantom{x}\hspace{12pt}var); sparse(repmat(1:N-2,n1+n2+2*n3,1),[indexM([type1; type2; type3],2:end-1); \\
			\phantom{x}\hspace{12pt}indexM(type3,1:end-2)],[ones(n1,N-2); ones(n2,N-2)/2; ones(2*n3,N-2)],N-2,var);\\
			sparse(1,[indexM(type1,end);indexM(type2,end);indexM(type3,end-1)],[ones(n1,1); \\
			\phantom{x}\hspace{12pt}ones(n2,1)/2; ones(n3,1)],1,var)];\\
			\\
			bineq = [bineq; ones(N,1)];\\
			\\
			{\color{Green}\% constraint on max weight (\ref{constraints}c)}\\
			\\
			Aineq = [Aineq;\\
			sparse(1,indexM([type1; type2],:),repmat(input([type1; type2],3),1,N),1,var)+...\\
			\phantom{x}\hspace{12pt}sparse(1,indexM(type3,1:end-1),repmat(input(type3,3),1,N-1),1,var)];\\
			\\
			bineq = [bineq; Wp];\\
			\\
			{\color{Green}\% constraint on the center of gravity (\ref{constraints}d)}\\
			\\
			Aineq = [Aineq;\\
			sparse(1,indexM([type1; type2],:),repmat(input([type1; type2],3),1,N).*(d([type1; \\
			\phantom{x}\hspace{12pt}type2],:)-xmax\_cg),1,var)+sparse(1,indexM(type3,1:end-1),repmat(input(type3,3),...\\
			\phantom{x}\hspace{12pt}1,N-1).*(d(type3,1:end-1)-xmax\_cg),1,var)];\\
			\\
			bineq = [bineq; -(xe\_cg-xmax\_cg)*We];\\
			\\
			Aineq = [Aineq;\\
			sparse(1,indexM([type1; type2],:),repmat(-input([type1; type2],3),1,N).*(d([type1;\\
			\phantom{x}\hspace{12pt}type2],:)-xmin\_cg),1,var)+sparse(1,indexM(type3,1:end-1),repmat(-input(type3,3),...\\
			\phantom{x}\hspace{12pt}1,N-1).*(d(type3,1:end-1)-xmin\_cg),1,var)];\\
			\\
			bineq = [bineq; (xe\_cg-xmin\_cg)*We];\\
			\\
			{\color{Green}\% constraint on the shear curve (\ref{constraints}e)}\\
			\\	
			{\color{Blu}for} j=1:floor(N/2)\\
			\phantom{x}\hspace{12pt}Aineq = [Aineq;\\
			\phantom{x}\hspace{12pt}sparse(1,indexM([type1; type2],1:j),repmat(input([type1; type2],3),1,j),1,var)+...\\
			\phantom{x}\hspace{12pt}sparse(1,indexM(type3,1:j),[repmat(input(type3,3),1,j-1) input(type3,3)/2],1,var)];\\
			\\
			\phantom{x}\hspace{12pt}bineq = [bineq; Smax\_0*j/floor(N/2)];\\
			\\
			\phantom{x}\hspace{12pt}Aineq = [Aineq;\\
			\phantom{x}\hspace{12pt}sparse(1,indexM([type1; type2],end:-1:end-j+1),repmat(input([type1;\\
			\phantom{x}\hspace{24pt}type2],3),1,j),1,var)+sparse(1,indexM(type3,end-1:-1:end-j),[repmat(input(type3,...\\
			\phantom{x}\hspace{24pt}3),1,j-1) input(type3,3)/2],1,var)];\\
			\\
			\phantom{x}\hspace{12pt}bineq = [bineq; Smax\_0*j/floor(N/2)];    \\
			{\color{Blu}end}\\
			\\
			{\color{Green}\% objective function (\ref{objective})}\\
			\\
			f = full(-sparse(1,indexM([type1; type2],:),repmat(input([type1; type2],3),1,N),1,...\\
			\phantom{x}\hspace{12pt}var)-sparse(1,indexM(type3,1:N-1),repmat(input(type3,3),1,N-1),1,var));
		}
	\end{minipage}
\end{figure*}
\begin{figure*}[htb]
	\begin{minipage}[t]{2\textwidth}
		{\fontfamily{qcr}\selectfont
			{\color{Green}\% generate output problem structure can be tested in Matlab using intlinprog}\\
			\\
			problem.Aeq = [];\\
			problem.beq = [];\\
			problem.Aineq = Aineq;\\
			problem.bineq = bineq;\\
			problem.f = f;\\
			problem.lb = zeros(var,1);\\
			problem.ub = ones(var,1);\\
			problem.solver = 'intlinprog';\\
			problem.options = [];\\
			problem.intcon = 1:var;\\
			\\
			{\color{Green}\% generate indexes to evaluate the center of gravity as \\
				\% x\_cg = (sparse(1,J,VL)*X+xe\_cg*We)./(sparse(1,J,V)*X+We)}\\	
			\\	
			J = indexM([type1; type2],:);\\
			dummy = indexM(type3,1:end-1);\\
			J = [J(:); dummy(:)];\\
			\\
			VL = repmat(input([type1; type2],3),1,N).*d([type1; type2],:);\\
			dummy = repmat(input(type3,3),1,N-1).*d(type3,1:end-1);\\
			VL = [VL(:); dummy(:)];\\
			\\
			V = repmat(input([type1; type2],3),1,N);\\
			dummy = repmat(input(type3,3),1,N-1);\\
			V = [V(:); dummy(:)];\\
			\\
			\\
			{\color{Blu}function} input = container\_distribution\_generator(n1,n2,n3,N)\\
			\\
			w1 = ([(3500-1500)/3*randn(n1*1000,1)+1500; (3500-1500)/3*randn(n1*1000,1)+3500]);\\
			w1 = round(w1((w1>1300)\&(w1<3700))*20/N);\\
			\\
			w2 = ([(1800-700)/3*randn(n2*1000,1)+700; (1800-700)/3*randn(n2*1000,1)+1800]);\\
			w2 = round(w2((w2>500)\&(w2<2000))*20/N);\\
			\\
			w3 = ([(7000-3200)/3*randn(n3*1000,1)+3200; (7000-3200)/3*randn(n3*1000,1)+7000]);\\
			w3 = round(w3((w3>3000)\&(w3<7200))*20/N);\\
			\\
			input = [(1:n1).'   ones(n1,1) w1(randperm(length(w1),n1));\\
			\phantom{x}\hspace{12pt}n1+(1:n2).' 2*ones(n2,1) w2(randperm(length(w2),n2));\\
			\phantom{x}\hspace{12pt}n1+n2+(1:n3).' 3*ones(n3,1) w3(randperm(length(w3),n3))];\\
		}	
	\end{minipage}
	\caption{Matlab code to generate Airbus ALO problems of any size. \label{matlabcode}}
\end{figure*}

\end{document}